\documentclass[12pt,preprint]{aastex}

\begin{document}

\title{Complex Variability of the H$\alpha$ Emission Line Profile of the T Tauri Binary System KH~15D:\\
The Influence of Orbital Phase, Occultation by the Circumbinary Disk, and Accretion Phenomenae} 

\author{Catrina M. Hamilton}
\affil{Department of Physics and Astronomy, Dickinson College, Carlisle, PA 
17013}
\email{hamiltoc@dickinson.edu}

\author{Christopher M. Johns-Krull}
\affil{Department of Physics and Astronomy, Rice University, Houston, TX
77005}
\email{cmj@rice.edu}

\author{Reinhard Mundt}
\affil{Max-Planck-Institut f\"{u}r Astronomie, K\"{o}nigstuhl 17,
D-69117 Heidelberg, Germany}
\email{mundt@mpia.de}

\author{William Herbst}
\affil{Astronomy Department, Wesleyan University, Middletown, CT
06459}
\email{wherbst@wesleyan.edu}

\author{Joshua N. Winn}
\affil{Department of Physics, Massachusetts Institute of Technology,
Cambridge, MA 02139}
\email{jwinn@mit.edu}

\begin{abstract}

We have obtained 48 high resolution echelle spectra of the pre-main sequence
eclipsing binary system KH~15D (V582 Mon, P = 48.37 d, $e$ $\sim$ 0.6, 
M$_{A}$ = 0.6 M$_{\odot}$, M$_{B}$ = 0.7 M$_{\odot}$).  The eclipses are caused by
a circumbinary disk seen nearly edge on, which at the epoch of these observations
completely obscured the orbit of star B and a large portion of the orbit of star A.  
The spectra were obtained over five contiguous observing  seasons from 
2001/2002 to 2005/2006 while star A was fully visible, fully 
occulted, and during several ingress and egress events.  The H$\alpha$ line 
profile shows dramatic changes in these time series data over timescales ranging from
days to years.  A fraction of the variations are due to ``edge effects" 
and depend only on the height of star A above or below the razor sharp edge of the occulting 
disk.  Other observed variations depend on the orbital phase: the H$\alpha$ 
emission line profile changes from an inverse P Cygni type profile during ingress
to an enhanced double-peaked profile, with both a blue and red emission component, 
during egress.  Each of these interpreted variations are complicated by the 
fact that there is also a chaotic, irregular component present in these profiles.  
We find that the complex data set can be largely understood in the 
context of accretion onto the stars from a circumbinary disk with gas flows 
as predicted by the models of eccentric T Tauri binaries put forward by 
Artymowicz \& Lubow, G\"{u}nther \& Kley, and de Val-Borro et al.  In particular, 
our data provide strong support for the pulsed accretion phenomenon, in which 
enhanced accretion occurs during and after perihelion passage.

\end{abstract}


DRAFT VERSION: 
\date{\today}

\section{Introduction}

KH~15D, first observed at Van Vleck Observatory in 1995, gained its notoriety due to 
the regular and rather large amplitude variability it displayed (Kearns \& Herbst 1998).  
At that time, the system's light was observed to diminish by nearly 3 magnitudes every 
48.4 days, and remain in this diminished or eclipsed state for approximately 16 days, 
with an unusual return to near-normal brightness close to mid-eclipse.  By 2000, the central 
reversal in the light curve had faded and the length of time that the star spent in the eclipsed 
state had grown slightly ($\sim$1 day).  Spectra taken of the object in and out of eclipse had 
shown that there was effectively no color change between states, implying that the source of 
obscuration was either an optically thick disk or rather large particles  (Hamilton et al. 2001).  
Several models were proposed at this time (Herbst et al. 2002;  Barge \& Viton 2003; 
Agol et al. 2004) that involved some sort of warp or swarm of particles in a circumstellar disk 
around a single star.

KH~15D was also discovered to be the source of a bipolar jet revealed in H$\alpha$ 
and [OI] (Hamilton et al. 2003) whose possible launching mechanism has been discussed
in detail by Mundt et al. (2010).  The system is associated with a shocked 
H$_{2}$ emission filament (Deming, Charbonneau, \& Harrington 2004; Tokunaga et al. 2004). 
During eclipse, the spectrum of KH~15D often exhibits extended wings of H$\alpha$ 
emission, up to several hundreds of km s$^{-1}$, characteristic of actively accreting classical 
T Tauri stars (CTTSs). Out of eclipse, the EW of H$\alpha$ is only a few angstroms, 
which would normally lead to a classification as a weak-lined T Tauri star (WTTS)
(Hamilton et al. 2003).  It is perhaps best described as a weakly accreting T Tauri star.
Its age, estimated from membership in NGC 2264, is $\sim$ 3 Myr.

In 2004, it was discovered that KH~15D is a binary system.  Johnson et al. (2004) 
conducted a radial velocity survey of the system over the course of two observing
seasons (2002/2003 and 2003/2004) during which KH~15D underwent significant radial
velocity variations consistent with a binary companion with an orbital period equal to that
of the photometric period of 48 days.  Almost simultaneously, two models were put forward
to explain the changing light curve as the result of the progressive occultation of a pre-main sequence binary orbit by a precessing circumbinary inner disk or ring (Winn et al. 2004; 
Chiang \& Murray-Clay 2004).  This type of model had the advantage of also being 
consistent with the historical light curves obtained from a study of archival photographic 
plates (Winn et al. 2003; Johnson \& Winn 2004).  Winn et al. (2006; hereafter W06) further 
refined this model by including a wealth of additional photometric data available from the
literature (Johnson et al. 2005; Maffei et al. 2005; Hamilton et al. 2005; Barsunova et al. 2005;
Kusakabe et al. 2005).

Our current working model of the system (W06) has the distinguishing aspect that the system 
is fortuitously observed nearly along the plane of an opaque circumbinary ring that 
is, itself, somewhat tilted with respect to the binary's orbital plane.  To orient the reader, 
a schematic of the KH~15D system is shown in Figure 1.  The ring is presumably the 
inner part ($<$ 5 AU) of a more extensive circumbinary disk (CBD) that provides gas for 
continuing accretion. The outer disk has recently been detected at mm wavelengths with
the SMA (Herbst \& Wilner, in preparation).  Precession of the ring has, over the past 
50 years, gradually occulted the orbit of the binary as projected on the sky.  Various studies 
have shown (e.g., Herbst et al. 2002; Herbst et al. 2010) that the edge of this occulting 
ring is surprisingly sharp (much less than one stellar radius) and behaves like a knife edge. 
For a more general discussion of the current properties of the KH~15D light curve and 
its interpretation, the reader is referred to Herbst et al. (2010).

When first noticed in 1995 (Kearns \& Herbst 1998) the occulting edge had just about 
completely covered the orbit of the more massive star (designated star B following the
nomenclature of W06). From 1995 to 2009 the knife edge moved progressively across the 
orbit of star A, producing the observed light curve (Hamilton et al. 2005; Herbst et al. 2010). 
During this time the system exhibited dramatic photometric variations on the orbital cycle 
as star A regularly rose and set with respect to the ring horizon. Throughout  part of this 
phase of its evolution (2001-2006) we were able to obtain a substantial number of spectra 
(N=48) of the system at a variety of orbital phases and heights of star A above and below 
the ring edge.  This has allowed us to do a kind of ``occultation mapping" of gas flows in 
the magnetosphere of star A and perhaps elsewhere within the inner cavity of the CBD. 
By far the most useful spectral feature available to us for this work is H$\alpha$. 

Hamilton et al. (2003) used 3 spectra obtained during one cycle in 2001 to infer the 
presence of a jet and a magnetospheric accretion flow.  At that time, the system was believed to
be only a single star with a circumstellar disk containing a warp.  In this paper, we present 
the analysis of additional high resolution spectroscopic data of the KH~15D system, specifically, the H$\alpha$ emission line profiles, obtained while in its bright state, during ingress/egress, 
and during eclipse over the course of a 5 year observing period (2001-2006).
Section 2 discusses the observations and reductions.  The H$\alpha$ emission
line profiles are presented and characterized in Section 3.  Profile decompositions based
on gaussian fitting and the velocity behavior of the absorption component are discussed in 
Section 4.  The high velocity emission component of the H$\alpha$ line profile
is presented in Section 5, and constraints on the size of the magnetosphere
of star A are given in Section 6.  General conclusions are drawn in Section 7. 

\section{Observations and Data Reduction}

The high resolution echelle spectra analyzed here were obtained during five 
observing seasons from 2001/2002 through 2005/2006 at various observing 
facilities.  Table 1 gives the UT dates of the observations, telescope/instrument used, 
wavelength coverage and resolution.  We attempted to exploit the  
``natural coronagraph" that this system represents by focusing primarily on phases 
during or near ingress and egress, however, spectra were also obtained near mid-eclipse,
and when star A was well out of eclipse. Figure 2 shows the Cousins $I$-band ($I_{C}$) 
data from Hamilton et al. (2005) or Herbst et al. (2010) phased with the orbital period for each 
of the five observing seasons overlain with the dates on which spectra were obtained. 
It is clear that the width of the eclipse increased significantly between the beginning and 
end of the data acquisition period so a particular orbital phase that represented an 
``out-of-eclipse" observation in 2001/2002 might be ``in eclipse" in 2005/2006.   This is 
demonstrated in Figure 3.  Each spectrum is characterized, therefore, by two numbers: 
orbital phase, ranging from --0.5 to +0.5 with 0.0 representing the time of mid-eclipse, 
and the position of star A with respect to the ring edge, ranging from $\Delta$X = --4 
(4 stellar radii above the edge) to $\Delta$X = +10 (10 stellar radii below the ring edge). 
The parameter $\Delta$X used here comes from Model 3 of W06\footnote{Model 3 employs
the astrophysical constraint that star A be less massive than star B because it is less 
luminous.  This constraint produces a model that is in best agreement with the observed
photometry through 2006.} and is the elevation of the center of star A above (--) or below 
(+) the edge of the occulting disk in units of the radius of star A 
(R$_{A}$ = 1.3 R$_{\odot}$ = 9 x 10$^{8}$ m).  

As is evident from Table 1, there were five different instruments employed to obtain the 
observational material analyzed here. These are: the Keck 10-m telescope and High 
Resolution Echelle Spectrometer (HIRES; solid red lines in Fig. 2), the European Southern 
Observatory's 8.2-m Very Large Telescope and UV-Visual Echelle Spectrograph 
(UVES; long-dashed blue lines), the 2.1-m Otto Struve Telescope at McDonald 
Observatory and Sandiford Cassegrain Echelle Spectrometer (CE; dash-dot green lines), 
the 8-m Hobby-Ebberly Telescope at McDonald Observatory and High Resolution 
Spectrograph (HRS; dashed pink lines) and the 6.5-m Magellan II (Clay) Telescope 
and Magellan Inamori Kyocera Echelle Spectrograph (MIKE; dash-dot-dot-dot orange lines). 
Brief descriptions of the data and reduction process are now given for each instrument.

\subsection{The VLT/UVES Spectra}

The data obtained with the VLT and UVES used in this analysis were
taken during the 2001/2002 and 2004/2005 observing seasons.  The data from
2001/2002 were obtained in queue mode and have already been published.  The
reader is referred to Hamilton et al. (2003) for more information regarding
their reduction.  The data taken during the 2004/2005 observing season
were reduced with a set of custom echelle reduction routines written in
IDL.  The data reduction procedure is described by Valenti (1994) and
Hinkle et al. (2000) and includes bias subtraction, flat fielding by a 
normalized flat spectrum, scattered light subtraction, and optimal extraction
of the spectrum.  Due to nebular H$\alpha$ emission near KH~15D, it is
important for this study to perform a sky subtraction when reducing
the spectra.  The VLT/UVES spectral format contains enough room between
the orders that background sky spectra are recorded both above and below the
stellar spectrum in the slit.  The spectral traces defining the
order locations for the orders in the stellar spectrum were offset in each
direction along the slit and sky spectra were extracted.  The two sky spectra
were averaged and used to subtract the sky background from the spectra of
KH~15D.  The wavelength solution was determined by fitting a 
two-dimensional polynomial to $n\lambda$ as function of pixel and order 
number, $n$, for several hundred extracted thorium lines observed from an 
internal lamp assembly.

\subsection{The Keck/HIRES Spectra}

Some of the data that were obtained with the Keck I telescope and HIRES
for this project were also used as a part of the radial velocity study 
Johnson et al. (2004).  A full description of the reduction 
procedure used for the Keck data obtained in 2003/2004 is given in that 
paper.  Additional Keck data obtained in Feb 2005 were reduced using the 
same custom IDL echelle reduction routines referenced above in \S{2.1}.  
Similar to the VLT/UVES data, sky subtraction was also performed for the
Keck/HIRES data.  

\subsection{The McDonald Observatory 2.1-m/CE Spectra}

The January 2004 data obtained with the McDonald Observatory 2.1-m Otto
Struve Telescope and CE spectrometer were 
also used as a part of the radial velocity survey presented by
Johnson et al. (2004).  The reader is again directed to that paper for a
full description of the reduction procedures for these spectra.

\subsubsection{2.1-m Sky Subtraction} 

The spectral format of the CE is quite compact, requiring a relatively
short slit (in this case 2.$^{''}$5) in order to keep the orders
well separated on the CCD.  As a result, traditional sky subtraction with
this system is not possible.  As mentioned above, there is an H$\alpha$ 
component in the line profiles of KH~15D due to the nebular emission from 
NGC 2264, the cluster in which KH~15D is located.  In order to study the
intrinsic variations of the H$\alpha$ profile of KH~15D, it is helpful
to remove this nebular component.  To do so, we utilized a sky spectrum 
obtained with one of the HET/HRS spectra of KH~15D (see below) as a proxy 
for sky subtracting the McDonald 2.1-m CE data.  After continuum normalizing
the H$\alpha$ order in the 2.1-m CE spectra, we scaled the HET sky 
spectrum so that when it is subtracted from the 2.1-m CE spectra the nearby
[NII] $\lambda$6583 emission line is entirely removed.  This line is a
pure nebular emission line, so the underlying assumption in this procedure
is that the ratio of this line to the nebular H$\alpha$ line stays 
constant throughout the nebula surrounding KH~15D.  As a check on this
assumption, we examined the spectra taken with Keck/HIRES which has a 
long enough slit to separately extract sky spectra on either side of the
stellar spectrum.  We scaled the sky spectrum on one side of the stellar
specrum so that the flux in the [NII] line matched that from the sky 
spectrum taken on the other side of the star.  We then differenced the
two sky spectra.  In each case, no residual H$\alpha$ flux is detected,
and the 3$\sigma$ upper limit on the H$\alpha$ flux in the difference
spectrum is $\sim 2$\% of the nebular H$\alpha$ emission.  It thus appears
that this is a fairly robust way to remove the sky$+$nebular emission 
from the 2.1-m CE spectra.

\subsection{The Magellan/MIKE Spectra}

The MIKE spectra were reduced with the ``MIKE redux" code written by S. 
Burles, X. Prochaska, and R. Bernsteini (see 
http://web.mit.edu/$\sim$burles/www/MIKE/).  This reduction package 
performs bias subtraction, flat fielding, order-edge tracing, sky subtraction, and optimal 
extraction.  Wavelength calibration is also performed as part of the package
using spectra of a Thorium-Argon arc lamp obtained just prior to each 
observation of KH~15D.  The wavelength solution for the MIKE data
was performed with the custom IDL software described above and referenced
in Valenti (1994) and Hinkle et al. (2000).

\subsection{The HET/HRS Spectra}

The HET/HRS spectra used here were all reduced with the custom IDL
software described above in \S{2.1} and referenced in 
Valenti (1994) and Hinkle et al. (2000).  The HRS instrument at the HET is a 
fiber fed spectrometer.  For each observation of KH~15D, sky fibers were placed on
either side of the star fiber.  As a result of the spectra from these
sky fibers appearing on the CCD, the optimal extraction routines in the
reduction package were not used.  Instead, the counts normal to the
dispersion direction were summed at each wavelength in order to produce
the extracted spectrum.  The observations used here come from two programs
looking at KH~15D.  As a result, about half the observations used 
2$^{\prime\prime}$ fibers while the other half used 3$^{\prime\prime}$ fibers.
The throughput of the fibers used to feed the stellar and sky spectra
vary somewhat.  As a result, the two sky spectra were summed to increase
the signal to noise and the result was then scaled to match the [NII] 
$\lambda$6583 flux in the stellar spectrum before the sky subtraction
was performed.  Spectra of a Thorium-Argon lamp were also extracted 
for each night and again a wavelength solution was created by fitting a
two-dimensional polynomial to $n\lambda$ as function of pixel and order
number, $n$, for several hundred thorium lines.

\section{The H$\alpha$ Emission Line Profiles}

In order to interpret the wealth of complex information produced by the occultation 
mapping of the H$\alpha$ line profile we required the height of star A above or below the 
occulting screen, which is taken from Model 3 of W06. Table 2 lists the Julian Date of 
observation, telescope/instrument, orbital phase, the height of Star A above the 
disk, the radial velocity (RV) of star A as predicted by W06, the radial velocity of star B
as predicted by W06, the calculated or measured (when available) Cousins $R$ 
magnitude ($R_{C}$), the measured $I_{C}$ (from Hamilton et al. 2005 
or Herbst et al. 2010), the flux ratio used to scale the normalized spectrum, and the 
barycentric correction.   

All profiles have been continuum-fit and normalized to 1.0.  Their flux was then 
scaled in reference to the 2002 December 10 out-of-eclipse spectrum obtained with
the HET at McDonald Observatory.  The 2002 December 10 spectrum was 
chosen as the reference spectrum to which all others would be scaled because star A
had the largest height above the occulting screen as predicted by the W06 model on 
that date and presumably represents the H$\alpha$ profile most intrinsic to star A.  
The scale factor was calculated by computing the difference between the 
$R_{C}$ magnitude on the particular date of interest and the $R_{C}$ magnitude 
observed on 2002 December 10.  
The flux on 2002 December 10 was assumed to be 1.00.  The Cousins $R$ filter 
encompasses the wavelength region corresponding to H$\alpha$, making it the 
best available proxy for measurement of the flux in the continuum near H$\alpha$.   
Actual observed $R_{C}$ magnitudes were used when available, however, the majority 
of the photometric monitoring during these seasons was only done in the Cousins $I$-band.  
When $R_{C}$ magnitudes were not available in the photometric database, they were
estimated from measured $I_{C}$ values using a color-magnitude relation from Hamilton
et al. (2005) based on USNO 2002/2003 data.  Typical photometric errors are $\sim$ 0.01 
out of eclipse and $\sim$ 0.1 in eclipse (Hamilton et al. 2005).

The scaled profiles are shown in Figures 4 and 5.  The profiles are plotted from left to right
in these figures according to distance above or below the projection of the edge of the 
occulting disk as calculated by W06.  They are shown, in reference to 
Figure 2, from out of eclipse, through ingress, during mid-eclipse, through egress, and 
out of eclipse again.  The profiles are plotted this way to assist in identifying whether or not 
there are systematic differences in the profiles given the specific location in the orbit, 
not just the height above or below the occulting disk.  All profiles have been corrected 
for the Earth's motion, and are shown in the rest frame of the center of mass of the system.   

\subsection{Characterizing the Variability}

When trying to understand the complex variations in the H$\alpha$ line profiles presented
here, one should consider three more or less independent factors influencing the variability:
(1) the ``edge effect" - occultation by the CBD as measured by the distance $\Delta$X with
respect to the occulting edge; (2) the ``orbital phase effect" - potential variability due to 
increased accretion on both stars or other changes during certain phases of the eccentric 
binary orbit (see de Val-Borro et al. 2011 and references therein); and (3) an irregular/chaotic 
component that is best seen when comparing the profiles taken at similar heights and 
orbital phase but at different times. We have several multi-day spectral sequences (some 
with coverage on consecutive days) obtained during ingress, egress,  and while the 
observable star is out of eclipse.  These observations occurred over the course of five 
different observing seasons so we can assess both the short-term and long-term variability 
of the H$\alpha$ emission line profile.  In the following section, we will begin by 
demonstrating the changes that are seen in the H$\alpha$ emission line profile as the 
stars move throughout their orbits, followed by changes that are produced as star A is
occulted by the CBD. Finally, we will discuss variations observed which can be attributed 
to the irregular nature of the T Tauri accretion process.

\subsubsection{Orbital Phase Variations and Occultation Variations}  

By looking at Figures 4 and 5, it is evident that the KH~15D system exhibits highly 
variable H$\alpha$ profiles.  We begin our analysis by taking the appropriately 
scaled (see \S 3) profiles from Figures 4 and 5 and averaging them according to the 
predicted distances of star A with respect to the edge of the occulting disk at the time of 
observation.  The spectra were split up into two sets of 7 different bins ($\Delta$X = --4 to --2, 
$\Delta$X = --2 to --1, $\Delta$X = --1 to 0, $\Delta$X = 0 to +1, $\Delta$X = +1 to +2, 
$\Delta$X = +2 to +4, and $\Delta$X = +7 to +10; see Table 3 for details), with the first set of 
bins for the spectra taken during ingress (approaching mid-eclipse, phases --0.5 to 0 
according to Figure 2), and the second set of bins for those taken during egress 
(following mid-eclipse, phases 0 to +0.5 according to Figure 2).  These sets of averaged 
profiles are displayed in Figure 6.  To orient the reader, ``time" in Figure 6 runs down the
left hand side of panels and then up the right hand side of panels back to the top.

As one examines the averaged, scaled emission line profiles presented 
in Figure 6, three distinct ``features" can be discerned and are indicated by arrows in
the appropriate panels.  They are:

\begin{enumerate}
\item The absorption component seen mostly out of eclipse and during ingress/egress.
\item The broad component mainly seen once star A is fully occulted.
\item The double peak emission profile primarily observable during full eclipse.
\end{enumerate}

\noindent
The top four panels in Figure 6 (a-d) represent the out-of-eclipse 
spectra obtained when $\Delta$X ranges from --4 to --1.  The next four panels (e-h) show 
the ingress and egress spectra obtained when $\Delta$X ranges from --1 to +1.  In these 
eight panels,  a ``central" absorption feature near velocity = 0 is predominant, and 
sometimes extends well below the stellar continuum.

The bottom six panels in Figure 6 (i-n) show the averaged H$\alpha$ profile after the star is 
completely obscured and at  a distance of $\Delta$X = +1 or greater.  Here, we see
a generally double-peaked emission line profile, with an underlying broad component.
The underlying broad component ranges in velocity from approximately -300 to +300 
km s$^{-1}$.  To emphasize the presence of this faint broad component, which is difficult 
to see in the regularly scaled plots, the profiles plotted in the bottom 6 panels of Figure 6 
have been multiplied by a factor of 10 and over-plotted on the same graph.  These 
modified profiles are shown as a dash-dot-dot-dot line in red.  Note that the bottom
two panels in Figure 6 (m and n) are identical.
  
The largest variation in the profiles examined here naturally occurs during the eclipse
itself as the star and its magnetosphere sink below, and alternatively rise 
from behind, the occulting screen.  To focus on these phases, we show in 
Figures 7 and 8, two ingress sequences; one from 2003 February and one from 
2004 December.  The ingresses are plotted two ways: 1) The top panel shows the 
fluxed profile with an offset, so that each individual profile is visible.  The zero point 
for each spectrum is indicated by the small horizontal line along the y-axis.  
2) The bottom panel shows each spectrum properly fluxed with its continuum 
adjusted back up to 1.0 so that they can be plotted on top of one another allowing one
to see the systematic decline of the flux in the far wings.

Figures 7 and 8 demonstrate that a ``typical" ingress spectrum may be characterized as an
inverse P Cygni type profile, consisting of red-shifted H$\alpha$ absorption, superimposed
upon blue-shifted H$\alpha$ emission.  We note that the absorption feature
starts out extending well below the stellar continuum.  As star A sinks below the occulting 
edge during ingress, the overall photospheric spectrum and all of the absorption that is 
seen in the H$\alpha$ profile drops to near zero flux levels.  
 
The profile that is left once the star is fully occulted and several stellar radii below 
the occulting edge primarily consists of a rather narrow double-peaked emission 
feature with peaks at $\sim$ $\pm$ 30 km s$^{-1}$, as well as an underlying faint broad 
component ranging in velocity from approximately -300 to 300 km s$^{-1}$, as mentioned
earlier.  This double-peaked emission feature was first noted by Hamilton et al. (2003), and 
its interpretation in terms of a bipolar jet has been discussed in great detail by 
Mundt et al. (2010), and will not be further discussed here.

After mid-eclipse, or periastron, star A emerges again, rising from behind the occulting
screen during egress.  During this phase, we see a more generally enhanced, roughly
symmetric, double-peaked H$\alpha$ emission profile with a slightly blue-shifted 
absorption feature, as shown in Figure 9.  Additionally, Figure 9 shows that during 
this particular egress, there is a reduction of peak emission flux as more of the star is 
exposed (i.e., 2005 February 28 vs. 2005 March 01).  This seems to be contrary to what 
one would expect; as more of the star is exposed, more of the H$\alpha$ emitting 
region, i.e., the magnetosphere, should be exposed, and therefore a greater amount of 
H$\alpha$ flux would be expected.  We return to this issue later.

In summary, the disappearance/reappearance of both the ``central" absorption feature 
as well as the overall photospheric spectrum observed in the KH~15D system can be 
attributed to the effect that the edge of the CBD and its occultation of star A and its close 
stellar environment has on the observed H$\alpha$ emission line spectrum.  Additionally,
the variation from inverse P Cygni profile to double-peaked profile must be dependent on 
where star A is in its orbit, since the inverse P Cygni profile is mostly observed out of eclipse
and into ingress.  The significance of these observations will be discussed in \S 5.

\subsubsection{Additional Variability}

T Tauri stars (TTSs) are generally characterized by their irregular variability due to 
accretion and other effects.  As a way of assessing the baseline variability of the visible 
TTS in this system, high-resolution spectra were obtained during the out-of-eclipse phases 
with the HET during the 2002/2003 (Hessman, PI) and 2005/2006 (Herbst, PI) observing 
seasons.  These data have been scaled by the appropriate flux ratio, had their
continuua adjusted back up to 1 for ease of comparison, and are shown in Figure 10.  
The left-hand side of Figure 10 shows the H$\alpha$ emission line profiles obtained 
out-of-eclipse in 2002 December and 2003 January/February.  The bottom panel shows the 
2002/2003 light curve and the vertical lines indicate when the spectra were obtained.  
The phases are listed in each plot as are the model predicted heights above the edge of 
the CBD on that given date.  On the right-hand side of Figure 10, we show the out-of-eclipse 
H$\alpha$ emission line profiles obtained during 2005 December and 2006 February.  
The bottom panel shows where the spectra were obtained in reference to the 2005/2006 
light curve.  

Focusing our attention first on the 2002/2003 season, we note that in this season,
the spectra were obtained when star A was located anywhere between $\sim $3 and 4
stellar radii above the edge of the CBD (see Figure 3 and Table 3).  The phases also range 
from +0.42 to --0.37, passing through phase $\pm $ 0.5.  We remind the reader that positive 
phases are those following mid-eclipse, representing the egress side of the orbit, 
while negative phases are those preceding mid-eclipse, representing the ingress side 
of the orbit.  A phase of $\pm $0.5 is exactly opposite mid-eclipse, which is denoted as 
phase = 0.

A comparison of the blue-shifted peaks observed in 2002 December and 2003 Jan/Feb 
does not show a great deal of variation in the maximum flux within the line.  In both cases,
the relative line flux is $\sim $ 2.1-2.2 at the peak.  The maximum flux occurs when the star is at its
greatest height ($\Delta$X = --4.02, phase = --0.48 for 2002 December, and 
$\Delta$X = --3.95, phase =  +0.49
for 2003 January).  What is highly variable, however, is the emission, and alternatively, the 
absorption, in the red-shifted portion of the H$\alpha$ line profile.  The interpretation of this
absorption feature and the cause of its variability is discussed in \S 4.

When comparing the observed profiles obtained during 2005/2006 season, one can see 
that the maximum flux in the blue-shifted peak in 2005 December is greater than that 
observed in 2006 February, one full orbital cycle apart, suggesting that this
may be the result of the intrinsic variability of TTSs.  However, it is 
important to note that the 2005 December spectra were taken as star A was
rising up from the edge of the CBD, as denoted by its X position (phases ranging from
0.38 to 0.46).  We see in this set of profiles that as the star continues to rise from behind 
the edge, the red-shifted emission increases in flux as the blue-shifted emission becomes 
somewhat narrower and lower in flux.  During this time, we also see an increase in the 
depth of the ``central" absorption feature, eventually extending below the continuum.

Comparatively, the profiles obtained in 2006 February (phases ranging from 0.39 to 0.5
and then to --0.47) are similar in peak blue-emission flux to the profiles shown on the 
left-hand side of Figure 10 from 2002/2003.  The maximum flux occurs around phase 
0.5 when the star is at its greatest predicted height above the occulting edge (2006 February 12).
Here we see both blue and red emission with a narrow ``central" absorption that extends 
below the stellar continuum for the first three profiles (2006 February 9, 10, and 12).  
As the star begins its ``descent" back toward the obscuring edge (phases have gone 
from + to --), the red-shifted emission feature disappears and is replaced by a broad absorption 
feature, apparently separate from the ``central" absorption, and centered around 
+100 km s$^{-1}$.  This type of change in profile shape is also seen in the top left panel 
of Figure 10 (2002 December).  As star A moves through phase 0.5 and begins its descent 
to ingress, the profile loses its red-ward emission as it is replaced by a broad 
red-shifted absorption feature.  Note also that the peak blue-emission flux in 2006 
February is similar in nature to the peak blue-emission flux seen in the out-of-eclipse 
spectra observed in 2002/2003, and shown on the left-hand side of Figure 10.  

In summary, the observations show that the H$\alpha$ emission line profile in
the KH~15D system varies systematically with the position of star A in its orbit and as a result of obscuration of the H$\alpha$ emitting region by the CBD.  We see a characteristic inverse 
P Cygni profile during ingress and a double-peaked, enhanced emission profile during egress.  
Our observations also reveal that the ``central" absorption feature is primarily red-shifted 
with respect to the systemic velocity during ingress and blue-shifted during egress.
Additionally, it appears as if the profile makes its transition from double-peak to inverse
P Cygni just as star A passes through phase = $\pm $ 0.5, or apastron.  In the next section, 
we explore the various components that make up the H$\alpha$ emission line profile.

\section{The Absorption Component}
 
It is clear from Figs. 4 and 5 that there is a substantial, but highly variable absorption 
component to the H$\alpha$ profiles when star A is above the occulting edge. Figure 6 
shows that this feature depends on the orbital phase of star A. During ingress the central 
absorption is broad and red-shifted relative to the systemic velocity (see panels a, c, e, and
g of Figure 6). During egress, the absorption feature tends to be narrower and more central or 
slightly blue-shifted (Figure 6, panels b, d, f, and h). Whenever star A is visible above the CBD, 
the absorption feature extends below the continuum, indicating that the gas responsible 
for it is at least partly projected onto the stellar photosphere.

To quantify the velocity behavior of the absorption component we have fit the profiles when 
the star is not fully eclipsed with three gaussians. Normally these were chosen to fit the red 
and blue emission wings and the central absorption. In one or two cases the red-shifted 
absorption was so strong that there was no blue emission wing visible and we fit two absorption 
components. Note that the intention here is to simply help quantify the mean velocity of 
the absorbing material clearly visible in the profiles. Examples of this fitting are shown 
in Figure 11, with the combined fit shown as the solid curve in red.  Note that in this figure, 
the left hand panels represents the ingress portion of the
orbit, while the right hand panels represents the egress portion of the orbit.  Additionally, we
draw attention to the fact that the top panels show out of eclipse spectra ($\Delta$X values 
$\sim$ +4), while the bottom panels represent spectra when the star was mostly eclipsed
($\Delta$X $\sim$ +0.5).  Plotting the profiles in this way helps to identify the differences
between ingress and egress.

There is a substantial amount of red-shifted absorption that comes and goes on the orbital 
cycle, being strongest during ingress and weaker or perhaps even absent during egress. 
It is not clear whether there is a separate narrow absorption component centered near the
systemic velocity or the velocity of star A. In at least one case shown in Figure 11, where 
we used two absorption components to model the line, this appears to be the case. The 
phase-dependent red-shifted component seen during ingress is, however, always present.

Figure 12 shows the velocity of the red-shifted absorption component plotted against binary 
phase. The top panel shows the velocity of the gas relative to the center of mass of the system,
while the bottom panel shows it in the frame of reference of star A.  Again, the feature is only
visible when at least some portion of star A is above the occulting edge, so there is a gap 
during perihelion passage. 
It is quite clear from the figure that as star A becomes visible on the far side of the disk 
hole from us, the central absorption is redshifted by only a small amount ($\sim$10 km s$^{-1}$)
with respect to star A.  But once the star ``turns the corner" and starts to move toward the
observer on its way to the near side of the gap where it will soon begin to be occulted, the 
highly red-shifted absorption component appears, with velocities reaching up to +100 
km s$^{-1}$ or more.

We believe that this relatively high velocity absorption component is a regular feature 
of the system that varies with orbital phase, and is not merely a stochastic feature 
attributable to unsteady accretion. The evidence is that the feature is present in every 
spectrum during the appropriate orbital phase, even those taken many cycles (and years) 
apart. It is not always precisely the same strength or velocity, but it is always identifiable 
as a feature of the H$\alpha$ profile. This suggests that it is a stable accretion feature 
(or ``stream") associated with star A. 

It is interesting that models of accretion in eccentric binary systems (e.g., de Val-Borro et al. 2011) 
predict precisely the sort of stable accretion streams that these observations require.  In these
models, the disk and binary are co-planar, however one can expect the general flow pattern
to be rather similar for a slightly inclined system (10-20$^{\circ}$).  Furthermore, one can
expect stronger accretion streams in the inclined system, which may help to understand 
why this WTTS has such strong accretion signatures.  
A schematic representation of how such an accretion stream would need to be configured to 
explain our observations was introduced in Figure 1.  Detailed modeling will be required to determine whether such a model satisfies the 
observations, but an accretion stream does seem promising as a way to understand this 
absorption component and its variation with orbital phase. 

Note that the maximum velocity observed, 100 km s$^{-1}$, is close to the free-fall velocity 
for gas falling from the inner edge of the CBD near 0.6 AU to $\sim$10 stellar radii. 
In this interpretation, then, we would identify 
the high velocity gas observed during ingress to be material in the accreting stream just 
outside the magnetosphere. It is quite possible that the gas accretes onto a disk at this 
location where it awaits a more rapid transfer to the star (or ejection into a jet) during 
perihelion passage; see next section). The lower velocity, but still accreting, gas seen 
during egress and perhaps at all times, may be part of a remnant disk of gas that did not 
accrete during perihelion passage or the outer parts of the accretion stream. 
Additionally, since the streams are in co-rotation, when star A is at apastron, gas will be seen
approaching with much larger velocities than the star, which would produce a blue-shifted 
wing.  Again, a detailed analysis of these events requires a dynamical model tuned to the 
properties of the KH~15D system and that is beyond the scope of this work.

\section{The High Velocity Emission Component}

Figure 6 shows clearly that the high velocity wings of the H-alpha profiles vary substantially 
and systematically with orbital phase. For single CTTSs, this emission is believed to arise 
within the magnetospheres of stars as they accrete material from a circumstellar disk 
and its strength is a proxy for the accretion rate.  During phases when star A is only partly 
occulted or entirely visible the high velocity component is associated with its magnetosphere. 
At other phases there may be a contribution from star B or from the 
combined magnetospheres. During these fully occulted phases much, if not all, of the light 
reaches us by reflection from the back wall of the CBD (Herbst et al. 2008), allowing us to observe 
phenomena occurring during perihelion passage that would otherwise be invisible.

The data show an interesting and clear trend that repeats in all cycles observed. Namely, we 
observe that the high velocity emission is more intense following perihelion passage (i.e. during 
egress) than it is prior to perihelion passage (i.e. during ingress). This may be seen by comparing 
the left hand side of Figure 6, corresponding to ingress, and the right hand side, corresponding 
to egress. The effects are particularly noticeable in panels near the bottom that apply to the fully 
occulted system just before and just after perihelion passage. On the left hand panels, one clearly 
sees the high velocity emission but it is nowhere near as intense as it is on the right hand panels. 
To make this clearer we have multiplied the relevant profiles by a factor of ten in the bottom six
panels.

Another way to demonstrate this point is to compare individual profiles obtained at the same 
locations with respect to the occulting disk and determine whether the ingress profiles look 
different from the egress profiles. This direct comparison is shown in Figure 13 where the spectra
have been properly fluxed and had their continuum adjusted back up to 1.0, and for two different 
values of $\Delta$X where we happened to have both ingress and egress spectra. It is clear that 
at both locations relative to the occulting disk (partly visible, $\Delta$X $\sim$ +0.5, top panel, and 
fully occulted, $\Delta$X = +1.09, bottom panel) the magnetospheric emission is more intense 
during egress than ingress. This will be quantified further in the next section where we discuss the 
equivalent widths of the blue emission wing in more detail.

We note that in addition to the phase-dependent variations discussed above there is 
clearly a time-dependent variation as well. It is not the case that every egress spectrum looks 
identical in its profile to others obtained at the same orbital phase. This time variability is illustrated 
in Figure 14 where we show some egress spectra obtained near the same orbital phase and at 
times when star A was fully occulted, or nearly so. These spectra are of somewhat lower resolution 
than many but serve to make the main point: there is substantial variation in the emission line 
wings that is independent of orbital phase or location relative to the occulting disk edge. The 2001 
Dec 20 spectrum is particularly notable, and was discussed in Hamilton et al. (2003) in the 
context of magnetospheric accretion. Here we see a very extended and substantial wing 
indicating that during this perihelion passage there was a substantial amount of activity.

In Figure 15 we show a sequence of spectra taken during a single egress while the system was 
fully occulted. Here, also, we see substantial variability in the high velocity emission, now on a 
daily timescale. There is some evidence for a high velocity ``feature" that shows up as a small 
bump in the blue wing on 2005 Dec 12 and migrates to lower velocity and higher intensity by 
2005 Dec 12. Whether this is related to the changing aspect of the star caused by stellar 
rotation and orbital motion or to some feature of the accretion is unknown. It is clear, however, 
that there is variability in the magnetospheric emission even on timescales as short as a day.

Our observations appear to confirm a general prediction of models of the binary accretion process 
in eccentric systems (Artymowicz \& Lubow 1996; G\"{u}nther \& Kley 2002; and de Val-Borro et al. 
2011) that accretion onto the stars occurs predominantly during periastron. It is during egress, i.e. 
after perihelion passage, that we see the most activity in the high velocity wings of the line and the 
greatest intensity overall. Similar results have been reported for DQ Tau (whose binary
parameters are very similar to those for KH~15D; see Table 4) by Mathieu et al. (1997) 
and Basri et al. (1997), for V4046~Sgr by Stempels \& Gahm (2004), for  UZ~Tau~E by 
Jensen et al. (2007), and for V773~Tau~A by Boden et al. (2007). 

In some cases mentioned above, the evidence for pulsed accretion includes phase-dependent 
continuum variations detected by broad band photometric studies. It has long been known that 
KH~15D shows a brightening near perihelion passage (see Hamilton et al. 2001; 
Hamilton et al. 2005; Herbst et al. 2010) at $I_{C}$ and some variations in color. 
The general phenomenon has been interpreted as 
the influence of star B, which is inferred to be the brighter of the pair, reaching its greatest 
extension toward the edge of the CBD at perihelion. At these orbital phases it, therefore, 
becomes the dominant source of the reflected light. It could, of course, be that some of this 
variation is caused by an increased brightness of one or both stars during perihelion passage. 
If the accretion is primarily onto the lower mass star (star A), we may not be able to witness the 
more dramatic brightness enhancements that occur, because star A is at its lowest point with 
respect to the disk edge then and perhaps completely invisible to us, even by reflected light.
This would, however, be in contrast to the current models, which predict that mass is 
preferentially channeled to the primary.  
Although the photometry cannot directly confirm the case for 
increased accretion during perihelion passage in the case of KH~15D, they are certainly 
not inconsistent with the spectroscopic evidence reported here.

\section{Constraints on the Size of the Magnetosphere of Star A}

Our data taken during eclipses can, at most, be used to constrain the size of the 
magnetosphere of star A in the vertical direction.  As outlined in previous papers 
(e.g., Herbst et al. 2002) the occulting edge of the circumbinary disk/ring is extremely 
``sharp," i.e.,  the transition region from optically thin to optically thick is less than 
$\sim$ 0.1 R$_{A}$ (and perhaps much less) so that the eclipses can be modeled 
with a ``knife-edge."  The magnetospheric emission model profiles of Hartmann and 
collaborators (Hartmann et al. 1994; Muzerolle et al. 1998; Muzerolle et al. 2001)  
show that the blue wing of these profiles is an emission component, while the red wing 
has contributions from both emission and absorption.  
We have calculated the emission flux and equivalent width of the blue-shifted 
wing of the H$\alpha$ profile between --280 km s$^{-1}$ and --85 km s$^{-1}$, ignoring
the velocity range from 0 to --85 km s$^{-1}$ because it is likely affected 
by emission from the jet.  For velocities less than -280 km s$^{-1}$, there is very little emission. 

In Figure 16  we show the fluxes (top panel) and equivalent width (lower panel) for two 
eclipse sequences obtained during ingress (2003 Feb and 2004 Dec; red squares
and blue triangles, respectively) and two eclipse sequences obtained during egress 
(2005 Feb and 2005 Dec; green stars and pink circles, respectively).  Each ingress or 
egress is indicated in the upper left corner of the figure, and the change in position from 
the center of mass of the binary of star A (r) is indicated for each ingress/egress shown.  
For reference, the binary semi-major axis is 0.25 AU = 41.4 R$_{A}$.  Note that for the 
ingress observations, time increases from left to right, while for the egress observations 
it is just the opposite.  To interpret the data displayed in Figure 16, we make use of the 
``knife-edge"  model, and assume that the emission region is spherically symmetric and 
homogeneously emitting as an initial thought experiment.  

We begin with the top panel of Figure 16, which shows the blue wing emission flux.  
Within the knife-edge model, the somewhat rapid decrease of the relative line fluxes during 
ingress can be most easily explained by a decreasing geometrical size of the emission 
region (i.e., less of the emission region is exposed as star A ``sets" below the edge of the 
occulting disk), provided this region is not highly variable on time scales of a few days.  
In principal, most of the flux data for the two egress phases can be explained in the same way. 
However, for the 2005 December egress, there is also a strong time variable component 
superimposed, which causes a strong decline in the line flux for the end of the egress phase 
(recall that for egress, increasing time runs from right to left).  As already outlined elsewhere in 
this paper (see \S 5) the line fluxes during egress are typically a factor 3 higher than during 
ingress, which has been attributed to increased accretion from the circumbinary disk/ring 
onto the two stars at the time near or after periastron passage.

To constrain the size of the H$\alpha$ emission region {\it relative to the size of 
the stellar disk}, we use the equivalent width (EW) of the emission contained within
the velocity range of --85 to --280 km s$^{-1}$, because the EW is a measure of the {\it ratio} 
between line flux and continuum flux.  To interpret the EW measurements within the 
context of the above mentioned ``knife-edge" model, assuming a spherically symmetric
emission line region, we consider two extreme cases: 1) the case of an emission 
region that is at least a few times larger than the stellar disk, and 2) the other extreme where 
the emission region is not extended and is roughly the same size as the stellar disk. 

In the first case, the EW should increase between $\Delta$X = 0 and $\Delta$X = +1, 
because at $\Delta$X = +1, all of the stellar disk is effectively covered and the only continuum 
emission visible would be that due to the scattered component.  If the emission region were
indeed larger than the stellar disk, say with a radius of 2R$_{A}$, only a fraction of that region
would be covered.  Therefore, the ratio between emission line flux and continuum flux should 
be highly increased.  We do, in fact, see a slight increase in the EW between $\Delta$X = 0 and 
$\Delta$X = +1 for both ingresses, however, it is more pronounced for the 2003 Feb ingress. 
If instead, we examine the second case where the emitting region is roughly the same size as
the stellar disk, we would not expect a change in the EW at all.  It is difficult to interpret the 
observed EW measurements in exactly this way, though the relatively small change in EW
suggests a relatively small magnetosphere.  
 
The egress data displayed in Figure 16 show that during egress the EW of
H$\alpha$ is about a factor of 2-4 greater than during ingress.  This means that
within the simplified spherically symmetric model considered here either
the radius or the surface brightness of the line emission region has increased. 
Unfortunately, the egress data are not well sampled between $\Delta$X = 0 and
$\Delta$X = +1 in order to check whether the radius of the emission line region has indeed
increased.  The 2005 Feb egress data suggest that there is a factor of 2 decrease in 
EW from $\Delta$X = +1 to +0.5, but this rests on only one data point.  In 2005 Dec, the
same sort of trend is seen (decreasing EW as the star moves closer to the occulting edge)
but the trend is not as steep for that egress, and the EW is on average greater.  

The EW data in Figure 16 can only constrain the extended emission perpendicular to the 
knife edge since it is believed that star A moves roughly up and down with respect to the
occulting edge.   We cannot exclude a disk-like emission region around the star with a 
disk of e.g. several stellar radii in size.  In fact, in the models of Hartmann et al. (1994), the 
vertical extent of their magnetospheric accretion flow is barely bigger than the star, however,
the magnetosphere extends out to 2-3 R$_{\star}$.   At most, we can argue that this disk like 
structure could be roughly 10-15\% thicker than the stellar diameter (in the direction 
perpendicular to the knife edge). 

If the H$\alpha$ emitting region were uniformly bright, we would expect that the EW 
measurements for both ingress and egress to be the same.  We show in Figure 17, the 
flux (top panel) and EW (bottom panel) of the H$\alpha$ emission contained within the 
velocity range of --85 to --280 km s$^{-1}$ for all observed profiles.  Open symbols represent 
ingress measurements, while filled symbols represent egress measurements.  Here we see a 
generally increasing EW for ingress from $\Delta$X = 0 to $\Delta$X = +1, and that all egress 
measurements are a factor of 2-4 times greater than ingress measurements.  

If we take the EW of the H$\alpha$ emitting region as a proxy for accretion rate, we can 
conclude that the accretion rate onto star A has increased significantly after periastron 
(phase = 0).  This has been modeled for eccentric binaries and is shown in Figure 2 of 
Artymowicz \& Lubow (1996), Figure 12 of G\"{u}nther \& Kley (2002), and
de Val-Borro et al. (2011).  Both G\"{u}nther \& Kley (2002) and de Val-Borro et al. (2011)
model the DQ Tau system, which can be used as a proxy for the KH~15D system (see Table
4). All models demonstrate that the accretion rates are greater during periastron passage,
however, both Artymowicz \& Lubow (1996) and G\"{u}nther \& Kley (2002) show that the
peak accretion occurs just prior to periastron passage.  

To compare our results to those presented in Figure 2 of Artymowicz \& Lubow (1996), we 
plot our EW measurements as a function of phase and approximated their model for
gas accretion rate as a simple gaussian.  Figure 18 shows that if we scale the gaussian to 
match our data on ingress (solid red line), we cannot match the egress accretion rate.  If 
instead, we scale the gaussian to match our egress data, we cannot match the ingress
observations (dashed blue line).  This suggests that the simulation does not quite accurately
predict the timing of the increase in accretion, or there may be something significant within the 
KH~15D system that is not included within the model, like a differential occultation of the 
magnetosphere.  The early calculations by Artymowicz \& Lubow (1996) used here 
were based on smooth particle hydrodynamics and rather coarse.  A new theoretical 
approach to this problem is warranted.


\section{Conclusions}

Our spectroscopic studies have revealed intrinsic variability and amazing structure in 
the H$\alpha$ line profiles caused by high velocity gas flows in the vicinity of the stars
within the KH~15D system.  The profiles observed during ingress and egress are 
distinctly different, changing from an inverse P-Cygni profile during ingress to an
enhanced double-peaked profile with broad extended wings during egress.  The
differences in these profiles can be understood in terms of models of accretion flows 
within an eccentric binary (Artymowicz \& Lubow 1996; G\"{u}nther \& Kley 2002; and 
de Val-Borro et al. 2011).  Measurements of the flux and EW of the blue-shifted emission 
located in the wings of H$\alpha$ (velocity ranging from --280 km s$^{-1}$ to --85 km s$^{-1}$)
give us an indication that the H$\alpha$ emitting region is compact and variable in brightness
on the time scale of the orbital period.  

There is potentially a great deal more information locked in these spectra than we have 
been able to extract. The present study will hopefully serve as an incentive and guide to 
more detailed modeling. The dynamics of the binary system are very well understood 
(W06) and it should be quite possible to model with definiteness the expected gas flows 
and accretion dynamics using codes that have already been applied to similar systems. 
These data could potentially serve as a critical test of such models. Some subtleties of the 
profile variations have undoubtedly escaped us and comparison with models of this 
particular system may reveal them.


\acknowledgments

We would like to thank the anonymous referee for the suggested improvements to this paper. 
We also thank Dr. Frederic V. Hessman for graciously allowing us to use the data he collected
with the HET.  C.M.H would like to thank Dr. Eric Mamajek and Dr. David James for helpful 
advice and discussions regarding the creation of Figure 1.  C.M.H. acknowledges partial 
support by the American Association of University Women through an American Fellowship. 
C.M.J.-K. acknowledges partial support by the NASA Origins
of Solar Systems program through the following grants to Rice
University: NNX08AH86G and NNX10AI53G.
W.H. acknowledges partial support by the NASA Origins
of Solar Systems program.
Some of the data presented herein were obtained at the W. M. Keck Observatory, which is 
operated as a scientific partnership among the California Institute of Technology, the 
University of California, and the National Aeronautics and Space Administration. The 
Observatory was made possible by the generous financial support of the W.M. Keck Foundation. 
The authors wish to recognize and acknowledge the very significant cultural role and reverence 
that the summit of Mauna Kea has always had within the indigenous Hawaiian community. 
We are most fortunate to have the opportunity to conduct observations from this mountain. 
The Hobby-Eberly Telescope (HET) is a joint project of the University of Texas at Austin, the 
Pennsylvania State University, Stanford University, Ludwig-Maximilians-Universit\"{a}t 
M\"{u}nchen, and Georg-August-Universit\"{a}t G\"{o}ttingen. The HET is named in honor of 
its principal benefactors, William P. Hobby and Robert E. Eberly. This paper is based in part on 
observations collected at the European Southern Observatory (Program 074.C- 0604A).  
C.M.H. would also like to thank Darla and Steve McKee, as well as the Masci family, for 
graciously supporting her and her daughter throughout the beginning, and final production, 
of this work, respectively.  This paper is dedicated to Robert Masci.


\clearpage
\begin{deluxetable}{lccc} 
\tabletypesize{\tiny}
\tablewidth{0pt}
\tablecaption{Spectroscopic Observations of KH~15D \label{tbl-1}} 
\tablehead{
\colhead{UT Date} & \colhead{Telescope/Instrument} & \colhead{Wavelength Coverage ($\mbox{\AA} $)} &
\colhead{${\lambda/\Delta\lambda}$} } 

\startdata

2001 Nov 29 & VLT/UVES   & 4800-6800 & 44,000 \\ 
2001 Dec 14 & VLT/UVES   & 4800-6800 & 44,000 \\
2001 Dec 20 & VLT/UVES   & 4800-6800 & 44,000 \\
2002 Dec 06 & HET/HRS    & 6380-7330\tablenotemark{a} & 15,000 \\ 
2002 Dec 10 & HET/HRS    & 6380-7330\tablenotemark{a} & 15,000 \\
2002 Dec 13 & HET/HRS    & 6380-7330\tablenotemark{a} & 15,000 \\
2003 Jan 23& HET/HRS    & 6380-7330\tablenotemark{a} & 15,000 \\
2003 Jan 26 & HET/HRS    & 6380-7330\tablenotemark{a} & 15,000 \\
2003 Feb 02 & HET/HRS    & 6380-7330\tablenotemark{a} & 15,000 \\
2003 Feb 05 & HET/HRS    & 6380-7330\tablenotemark{a} & 15,000 \\
2003 Feb 08 & Keck/HIRES & 4200-6600 & 70,000 \\ 
2003 Feb 09 & Keck/HIRES & 4200-6600 & 70,000 \\
2003 Feb 10 & Keck/HIRES & 4200-6600 & 70,000 \\
2003 Mar 06 & HET/HRS    & 6380-7330\tablenotemark{a} & 15,000 \\
2003 Mar 23 & HET/HRS    & 6380-7330\tablenotemark{a} & 15,000 \\
2003 Dec 16 & Keck/HIRES & 4400-6800 & 40,000 \\
2003 Dec 18 & Keck/HIRES & 4400-6800 & 40,000 \\
2004 Jan 04 & McD/CE     & 5600-6900 & 30,000 \\ 
2004 Jan 05 & McD/CE     & 5600-6900 & 30,000 \\
2004 Jan 10 & McD/CE     & 5600-6900 & 30,000 \\
2004 Feb 05 & Keck/HIRES & 4400-6800 & 39,000 \\ 
2004 Mar 10 & Keck/HIRES & 4700-7100 & 40,000 \\
2004 Mar 12 & Keck/HIRES & 4800-7100 & 40,000 \\
2004 Dec 10 & McD/CE     & 5600-6900 & 30,000 \\ 
2004 Dec 13 & VLT/UVES   & 5800-7300 & 44,000 \\ 
2004 Dec 14 & VLT/UVES   & 5900-7300 & 44,000 \\
2004 Dec 15 & VLT/UVES   & 5800-7300 & 44,000 \\
2004 Dec 16 & VLT/UVES   & 5800-7300 & 55,000 \\
2004 Dec 17 & VLT/UVES   & 5800-7300 & 55,000 \\
2004 Dec 18 & VLT/UVES   & 5800-7300 & 55,000 \\
2005 Feb 27 & Keck/HIRES & 4800-7100 & 42,000 \\
2005 Feb 28 & Keck/HIRES & 4800-7100 & 42,000 \\
2005 Mar 01 & Keck/HIRES & 4800-7100 & 42,000 \\
2005 Dec 12 & Mag/MIKE   & 5800-6800 & 25,000 \\
2005 Dec 13 & Mag/MIKE   & 5800-6800 & 25,000 \\
2005 Dec 14 & Mag/MIKE   & 5800-6800 & 25,000 \\
2005 Dec 15 & Mag/MIKE   & 5800-6800 & 25,000 \\
2005 Dec 20 & HET/HRS    & 5880-6770\tablenotemark{b} & 30,000 \\
2005 Dec 21 & HET/HRS    & 5880-6770\tablenotemark{b} & 30,000 \\
2005 Dec 22 & HET/HRS    & 5880-6770\tablenotemark{b} & 30,000 \\
2005 Dec 23 & HET/HRS    & 5880-6770\tablenotemark{b} & 30,000 \\
2005 Dec 24 & HET/HRS    & 5880-6770\tablenotemark{b} & 30,000 \\
2006 Feb 07 & HET/HRS    & 5880-6770\tablenotemark{b} & 30,000 \\
2006 Feb 09 & HET/HRS    & 5880-6770\tablenotemark{b} & 30,000 \\
2006 Feb 10 & HET/HRS    & 5880-6770\tablenotemark{b} & 30,000 \\
2006 Feb 12 & HET/HRS    & 5880-6770\tablenotemark{b} & 30,000 \\
2006 Feb 13 & HET/HRS    & 5880-6770\tablenotemark{b} & 30,000 \\
2006 Feb 14 & HET/HRS    & 5880-6770\tablenotemark{b} & 30,000 \\

\enddata

\tablenotetext{a}{These wavelengths were covered on the blue chip. In addition, the wavelength 
region 7480-8320 $\mbox{\AA} $ was covered on the red chip.}
\tablenotetext{b}{These wavelengths were covered on the blue chip. In addition, the wavelength 
region 6880-7800 $\mbox{\AA} $ was covered on the red chip.}

\end{deluxetable}

\clearpage
\begin{deluxetable}{lccccccccc}
\rotate  
\tabletypesize{\tiny}
\tablewidth{0pt}
\tablecaption{Parameters Used to Plot and Scale the H$\alpha$ Emission Line Profiles Shown in Figures 4 and 5. \label{tbl-2}}
\tablehead{
\colhead{Julian Date} & \colhead{Telescope/Instrument} & \colhead{Orbital Phase\tablenotemark{a}} & 
\colhead{Position of Star A\tablenotemark{b}} & 
\colhead{$RV_{A}$\tablenotemark{b}} & 
\colhead{$RV_{B}$\tablenotemark{b}} & \colhead{$R$\tablenotemark{c}} & 
\colhead{$I$\tablenotemark{d}} & \colhead{Flux Ratio\tablenotemark{e}} & 
\colhead{Barycentric Correction}  \\
\colhead{} & \colhead{} & \colhead{} & \colhead{} & \colhead{km s$^{-1}$} &  \colhead{km s$^{-1}$} & 
\colhead{} & \colhead{} & \colhead{} &  \colhead{km s$^{-1}$} 
}

\startdata

2452242.7446 & VLT/UVES   & -0.26 & -1.46 &  -9.95 &   4.52 & 15.28 & 14.49 & 0.990 &  16.23 \\
2452257.8271 & VLT/UVES   &  0.05 &  8.74 &  59.86 & -27.21 & 18.86 & 18.18 & 0.036 &   9.10 \\
2452263.7026 & VLT/UVES   &  0.17 &  1.95 &  -2.19 &   0.99 & 18.31 & 17.61 & 0.060 &   6.35 \\
2452614.9333 & HET/HRS    &  0.44 & -3.90 & -18.89 &   8.58 & 15.25 & 14.48 & 1.020 &  12.80 \\
2452618.9292 & HET/HRS    & -0.48 & -4.02 & -18.43 &   8.61 & 15.27 & 14.50 & 1.000 &  10.89 \\
2452621.7431 & HET/HRS    & -0.43 & -3.65 & -17.34 &   7.88 & 15.26 & 14.49 & 1.010 &   9.90 \\
2452662.6451 & HET/HRS    &  0.42 & -3.68 & -18.83 &   8.55 & 15.25 & 14.48 & 1.020 & -10.68 \\
2452665.8014 & HET/HRS    &  0.49 & -3.95 & -18.76 &   8.52 & 15.25 & 14.48 & 1.020 & -12.51 \\
2452672.6090 & HET/HRS    & -0.37 & -2.91 & -15.82 &   7.18 & 15.29 & 14.52 & 0.980 & -15.23 \\
2452675.7688 & HET/HRS    & -0.31 & -1.71 & -12.99 &   5.90 & 15.29 & 14.52 & 0.980 & -16.92 \\
2452678.8193 & Keck/HIRES & -0.26 & -0.12 &  -8.94 &   4.06 & 15.57 & 14.80 & 0.760 & -18.29 \\
2452679.8276 & Keck/HIRES & -0.24 &  0.50 &  -7.20 &   3.27 & 16.20 & 15.46 & 0.420 & -18.71 \\
2452680.8356 & Keck/HIRES & -0.22 &  1.17 &  -5.15 &   2.42 & 18.09 & 17.39 & 0.070 & -19.13 \\
2452704.6771 & HET/HRS    &  0.29 & -1.39 & -15.35 &   6.97 & 15.33 & 14.56 & 0.950 & -26.23 \\
2452721.6361 & HET/HRS    & -0.36 & -2.59 & -15.33 &   6.96 & 15.28 & 14.46 & 0.990 & -28.70 \\
2452990.0431 & Keck/HIRES &  0.18 &  2.72 &  -4.83 &   2.19 & 18.29$^{\ast}$ & 17.62 & 0.060 &   7.84 \\
2452992.0413 & Keck/HIRES &  0.22 &  1.09 & -10.63 &   4.83 & 18.11$^{\ast}$ & 17.42 & 0.070 &   6.84 \\
2453008.7960 & McD/CE     & -0.42 & -2.71 & -17.33 &   7.87 & 15.25 & 14.47 & 1.020 &  -1.33 \\
2453009.8729 & McD/CE     & -0.40 & -2.48 & -16.74 &   7.60 & 15.22 & 14.44 & 1.050 &  -2.06 \\
2453014.7920 & McD/CE     & -0.30 & -0.77 & -12.58 &   5.71 & 15.34 & 14.57 & 0.940 &  -4.42 \\
2453040.8025 & Keck/HIRES &  0.22 &  0.92 & -11.45 &   5.20 & 18.24 & 17.56 & 0.060 & -16.88 \\
2453074.7882 & Keck/HIRES & -0.07 &  7.61 &  23.23 & -10.56 & 18.85$^{\ast}$ & 18.25 & 0.040 & -27.41 \\
2453076.8125 & Keck/HIRES & -0.03 &  9.36 &  42.73 & -19.43 & 18.51$^{\ast}$ & 17.81 & 0.050 & -27.79 \\
2453349.8684 & McD/CE     & -0.38 & -1.31 & -15.86 &   7.20 & 15.28 & 14.51 & 0.990 &  10.85 \\
2453352.7976 & VLT/UVES   & -0.31 & -0.28 & -13.29 &   6.03 & 15.35 & 14.59 & 0.930 &   9.56 \\
2453353.6774 & VLT/UVES   & -0.29 &  0.11 & -12.31 &   5.59 & 15.51 & 14.74 & 0.800 &   9.28 \\
2453354.7199 & VLT/UVES   & -0.27 &  0.60 & -11.01 &   5.00 & 16.35 & 15.60 & 0.370 &   8.73 \\
2453355.6510 & VLT/UVES   & -0.25 &  1.09 &  -9.69 &   4.40 & 17.55 & 16.85 & 0.120 &   8.32 \\
2453356.6724 & VLT/UVES   & -0.23 &  1.66 &  -8.04 &   3.65 & 17.89 & 17.20 & 0.090 &   7.80 \\
2453357.6852 & VLT/UVES   & -0.21 &  2.28 &  -6.16 &   2.79 & 18.20 & 17.50 & 0.067 &   7.28 \\
2453428.8268 & Keck/HIRES &  0.26 &  1.09 & -13.25 &   6.02 & 17.31 & 16.60 & 0.150 & -24.91 \\
2453429.8250 & Keck/HIRES &  0.28 &  0.52 & -14.70 &   6.67 & 15.86 & 15.10 & 0.580 & -25.18 \\
2453430.8201 & Keck/HIRES &  0.30 &  0.02 & -15.84 &   7.19 & 15.47 & 14.70 & 0.830 & -25.45 \\
2453716.7708 & Mag/MIKE   &  0.21 &  3.31 &  -8.20 &   3.72 & 18.80 & 18.13 & 0.040 &   9.98 \\
2453717.7708 & Mag/MIKE   &  0.23 &  2.58 & -10.71 &   4.86 & 18.27 & 17.59 & 0.060 &   9.49 \\
2453718.7708 & Mag/MIKE   &  0.25 &  1.93 & -12.69 &   5.76 & 18.14 & 17.45 & 0.070 &   9.00 \\
2453719.7708 & Mag/MIKE   &  0.27 &  1.35 & -14.25 &   6.47 & 17.60 & 16.90 & 0.120 &   8.51 \\
2453724.7264 & HET/HRS    &  0.38 & -0.63 & -18.26 &   8.30 & 15.38 & 14.61 & 0.900 &   6.35 \\
2453725.7340 & HET/HRS    &  0.40 & -0.87 & -18.59 &   8.44 & 15.31 & 14.52 & 0.960 &   5.83 \\
2453726.7222 & HET/HRS    &  0.42 & -1.05 & -18.79 &   8.53 & 15.28 & 14.51 & 0.990 &   5.35 \\
2453727.7181 & HET/HRS    &  0.44 & -1.19 & -18.89 &   8.58 & 15.27 & 14.50 & 1.000 &   4.84 \\
2453728.7090 & HET/HRS    &  0.46 & -1.28 & -18.90 &   8.58 & 15.22 & 14.45 & 1.050 &   4.35 \\
2453773.7556 & HET/HRS    &  0.39 & -0.67 & -18.49 &   8.40 & 15.49 & 14.73 & 0.820 & -17.81 \\
2453775.7458 & HET/HRS    &  0.43 & -1.03 & -18.86 &   8.57 & 15.34 & 14.57 & 0.940 & -18.59 \\
2453776.5862 & HET/HRS    &  0.45 & -1.12 & -18.91 &   8.59 & 15.28 & 14.50 & 0.990 & -18.59 \\
2453778.7409 & HET/HRS    &  0.50 & -1.21 & -18.72 &   8.51 & 15.23 & 14.46 & 1.040 & -19.74 \\
2453779.5889 & HET/HRS    & -0.49 & -1.19 & -18.55 &   8.43 & 15.25 & 14.48 & 1.020 & -19.74 \\
2453780.7305 & HET/HRS    & -0.47 & -1.12 & -18.22 &   8.28 & 15.26 & 14.49 & 1.010 & -20.45 \\

\enddata

\tablenotetext{a}{The orbital phase was calculated using the following ephemeris for mideclipse: JD (mideclipse) = 2,453,077.59 + 48.37$E$.} 
\tablenotetext{b}{These values come from Model 3 of Winn et al. (2006).  The height above the disk
is given in units of the radius of star A.  The model predicted systemic velocity of 18.676 km s$^{-1}$
has been subtracted out of the radial velocities to put them in the reference frame of the system.} 
\tablenotetext{c}{The $R$ magnitudes were either measured directly (marked with an asterisk) or
calculated using the following relation derived from the 2002/2003 color data from
Hamilton et al. (2005): $R$-$I$ = 1.18301 - 0.0283473*$I$.}
\tablenotetext{d}{The $I$ magnitudes were taken from Hamilton et al. (2005) except for the 
2005/2006 observing season, which were taken from Herbst et al. (2010).}
\tablenotetext{e}{The flux ratios were calculated relative to the 2002 December 10
(JD = 2452721.6361) observation because it has the highest calculated position above the disk as
predicted by Model 3 of Winn et al. (2006), and therefore, is presumably the least obscured 
observation.}

\end{deluxetable}

\clearpage
\begin{deluxetable}{cccc} 
\rotate
\tabletypesize{\tiny}
\tablecaption{Dates and predicted heights for spectra included in each distance bin presented
in Figure 6. \label{tbl-3}}
\tablewidth{0pt}
\tablehead{
\colhead{Distance Bin} & \colhead{UT Date} & \colhead{Height (R$_{\star}$)} & \colhead{Obs/Instr} }

\startdata 

$\Delta$X = -4 to -2 & 2002 Dec 10 & -4.02 & HET/HRS \\
    (ingress)                & 2002 Dec 13 & -3.65 & HET/HRS \\
                                    & 2003 Feb 02 & -2.91 & HET/HRS \\
                                    & 2004 Jan 04 & -2.71 & McD/CE \\
                                    & 2003 Mar 23 & -2.59 & HET/HRS \\
                                    & 2004 Jan 05 & -2.48 & McD/CE \\

\hline                                    

$\Delta$X = -2 to -1 & 2003 Feb 05 & -1.71 & HET/HRS \\
    (ingress)                & 2001 Nov 29 & -1.46 & VLT/UVES \\
                                    & 2004 Dec 10 & -1.31 & McD/CE \\
                                    & 2006 Feb 13 & -1.19 & HET/HRS \\
                                    & 2006 Feb 14 & -1.12 & HET/HRS \\
                                    
\hline

$\Delta$X = -1 to 0 & 2004 Jan 10 & -0.77 & McD/CE \\
    (ingress)               & 2004 Dec 13 & -0.28 & VLT/UVES \\
                                   & 2003 Feb 08 & -0.12 & Keck/HIRES \\

\hline
                                   
$\Delta$X = 0 to 1 & 2004 Dec 14 & 0.11 & VLT/UVES \\
    (ingress)             & 2003 Feb 09 & 0.50 & Keck/HIRES \\
                                 & 2004 Dec 15 & 0.60 & VLT/UVES \\

\hline
                                 
$\Delta$X = 1 to 2 & 2004 Dec 16 & 1.09 & VLT/UVES \\
    (ingress)             & 2003 Feb 10 & 1.17 & Keck/HIRES \\
                                 & 2004 Dec 17 & 1.66 & VLT/UVES \\

\hline
                                 
$\Delta$X = 2 to 4 & 2004 Dec 18 & 2.28 & VLT/UVES \\

\hline

$\Delta$X = 7 to 10 & 2004 Mar 10 & 7.61 & Keck/HIRES \\
    (mid-eclipse)        & 2001 Dec 14 & 8.74 & VLT/UVES \\
                                    & 2004 Mar 12 & 9.36 & Keck/HIRES \\

\hline                                    

$\Delta$X = 4 to 2 & 2005 Dec 12 & 3.31 & Magellan/MIKE \\
       (egress)           & 2003 Dec 16 & 2.72 & Keck/HIRES \\
                                 & 2005 Dec 13 & 2.58 & Magellan/MIKE \\

\hline
                                                   
$\Delta$X = 2 to 1 & 2001 Dec 20 & 1.95 & VLT/UVES \\
       (egress)           & 2005 Dec 14 & 1.93 & Magellan/MIKE \\
                                 & 2005 Dec 15 & 1.35 & Magellan/MIKE \\
                                 & 2005 Feb 27 & 1.10 & Keck/HIRES \\
                                 & 2003 Dec 18 & 1.09 & Keck/HIRES \\

\hline
                                 
$\Delta$X = 1 to 0 & 2004 Feb 05 & 0.92 & Keck/HIRES \\
       (egress)           & 2005 Feb 28 & 0.52 & Keck/HIRES \\
                                 & 2005 Mar 01 & 0.02 & Keck/HIRES \\

\hline
                                 
$\Delta$X = 0 to -1 & 2005 Dec 20 & -0.63 & HET/HRS \\
       (egress)             & 2006 Feb 07 & -0.67 & HET/HRS \\
                                   & 2005 Dec 21 & -0.87 & HET/HRS \\

\hline
                                   
$\Delta$X = -1 to -2 & 2006 Feb 09 & -1.03 & HET/HRS \\
       (egress)              & 2005 Dec 22 & -1.05 & HET/HRS \\
                                    & 2006 Feb 10 & -1.12 & HET/HRS \\
                                    & 2005 Dec 23 & -1.19 & HET/HRS \\
                                    & 2006 Feb 12 & -1.21 & HET/HRS \\
                                    & 2005 Dec 24 & -1.28 & HET/HRS \\
                                    & 2003 Mar 06 & -1.39 & HET/HRS \\

\hline
                                    
$\Delta$X = -2 to -4 & 2003 Jan 23 & -3.68 & HET/HRS \\
       (egress)              & 2002 Dec 06 & -3.90 & HET/HRS \\
                                    & 2003 Jan 26& -3.95 & HET/HRS \\

\hline                                    

\enddata 

\end{deluxetable}

\clearpage
\begin{deluxetable}{cccccc} 
\tablecaption{Orbital Parameters for DQ~Tau and KH~15D. \label{tbl-3}}
\tablewidth{0pt}
\tablehead{
\colhead{} & \colhead{Period} & \colhead{} & \colhead{} & \colhead{$a$} & \colhead{}\\
\colhead{Binary System} & \colhead{(days)} & \colhead{$M_{2}/M_{1}$} & \colhead{$e$} 
& \colhead{(AU)} & \colhead{$i$} }

\startdata 

DQ~Tau\tablenotemark{a} & 15.804 & 0.97 & 0.56 & 0.13 & 23$^{\circ}$\\
KH~15D\tablenotemark{b} & 48.38 & 0.83 & 0.57 & 0.25 & 92.5$^{\circ}$\\

\hline

\enddata

\tablenotetext{a}{Parameters for DQ~Tau come from Mathieu et al. 1997.} 
\tablenotetext{b}{Parameters for KH~15D come from Model 3 of W06.} 

\end{deluxetable}

\clearpage
\begin{figure}
\centering
\includegraphics[width=6.5in,height=2.5in]{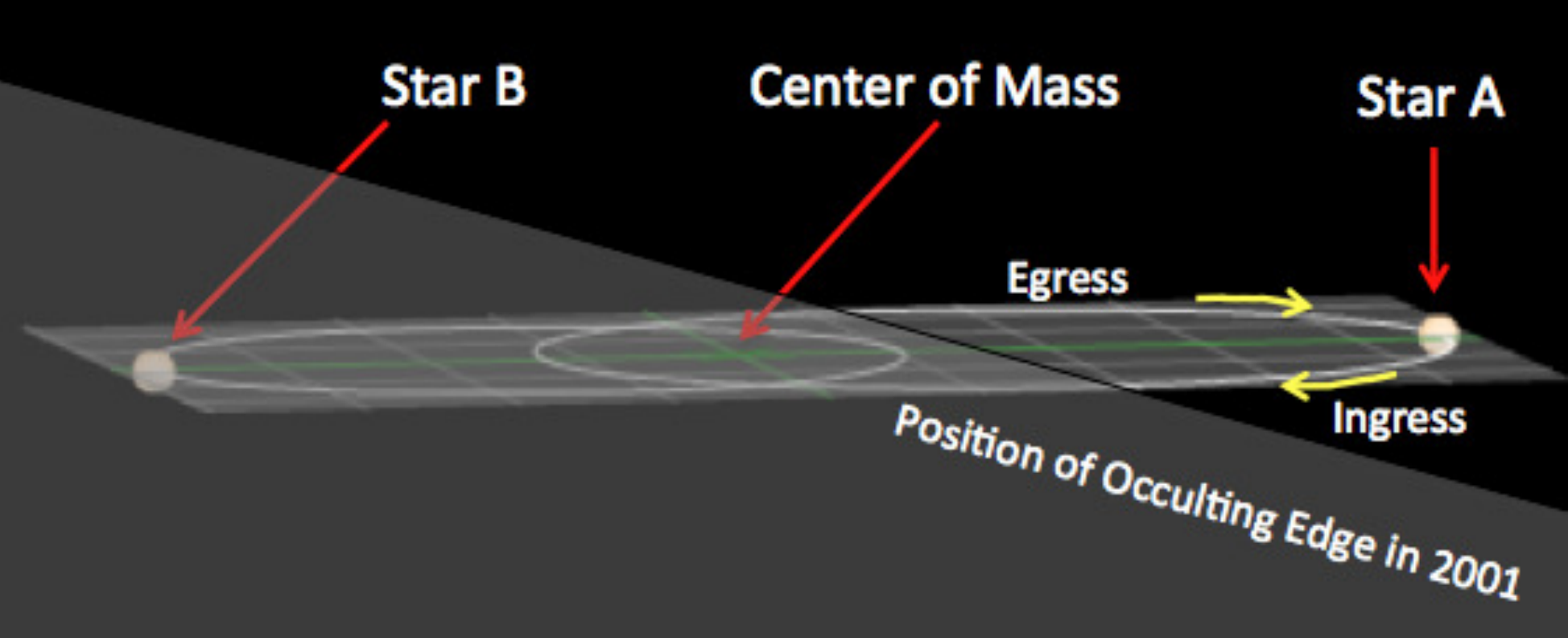}
\includegraphics[width=6.5in,height=4in]{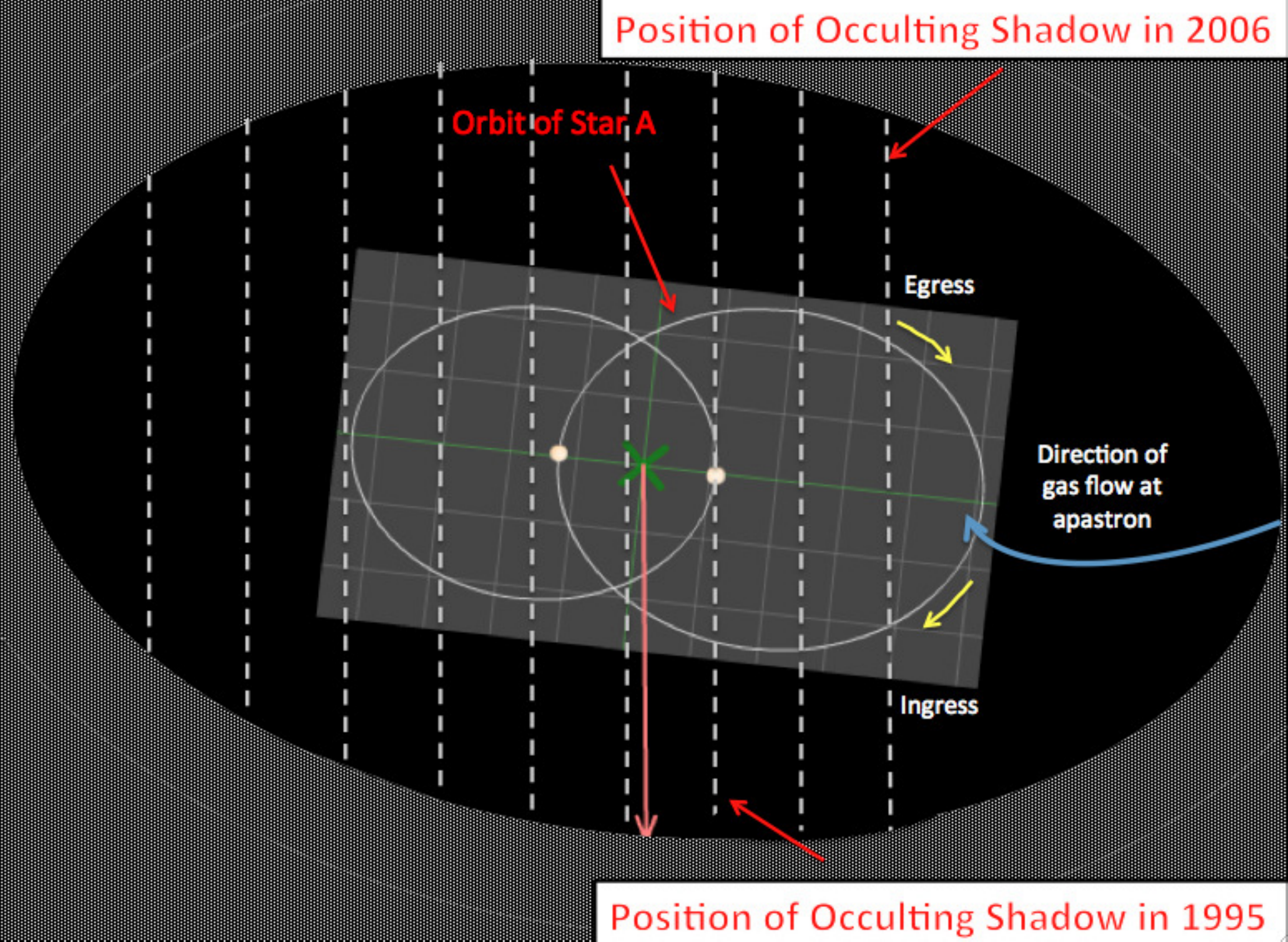}
\figcaption{The top panel depicts the KH~15D system roughly edge on with the
stars positioned at apastron. The bottom panel shows a top view of the system
with the stars positioned at periastron.  The pink arrow denotes the direction to Earth.  
The blue arrow to the right of the figure shows the inferred direction of the gas flow, 
as discussed in \S 4. {\footnotesize This figure was created with the help
of an eclipsing binary simulator produced by the University of Nebraska-Lincoln, 
freely available at http://astro.unl.edu/.}
\label{Fig. 1}}
\end{figure}

\clearpage
\begin{figure}
\centering
\includegraphics[totalheight=0.8\textheight]{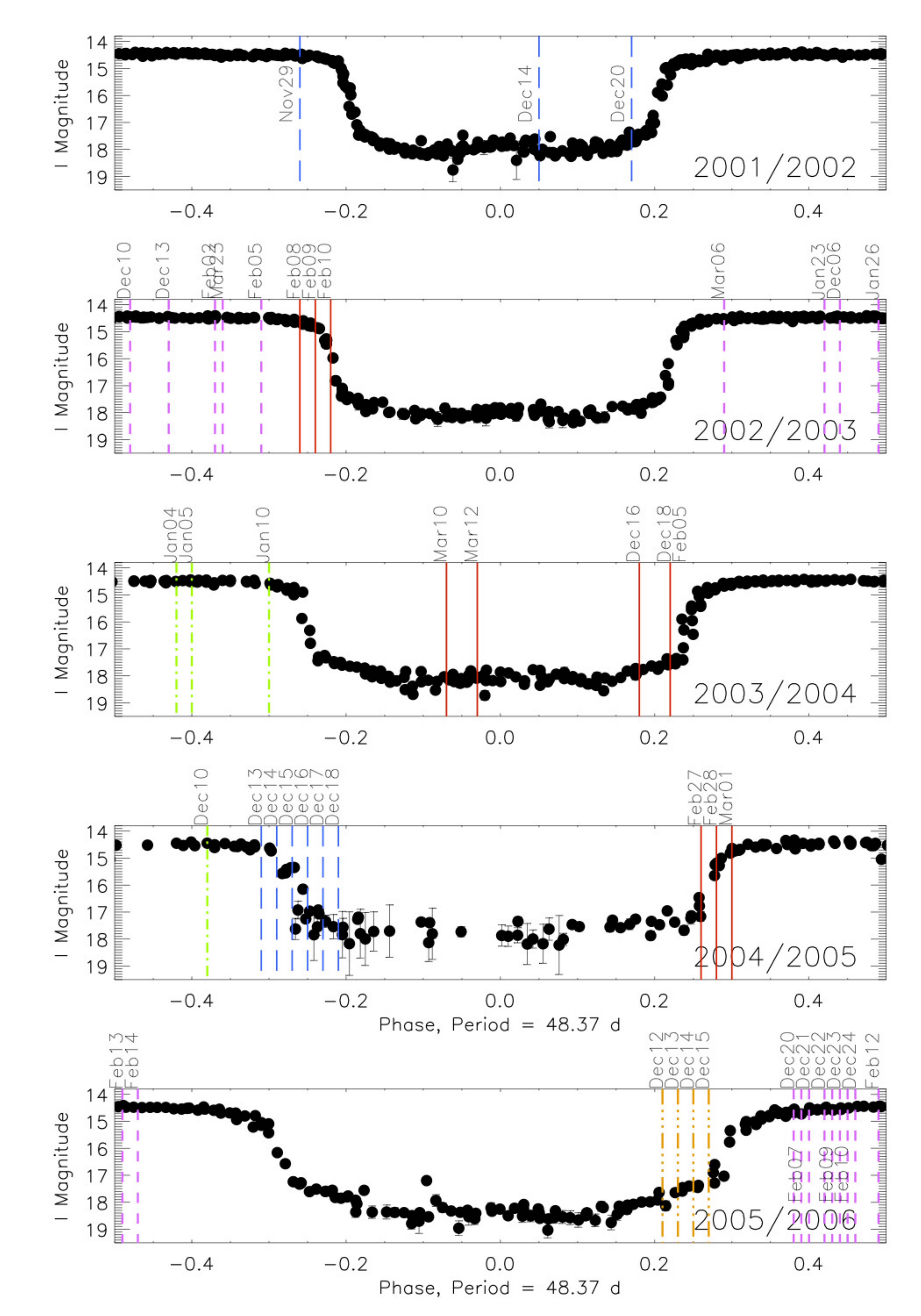}
\figcaption{Phases at which high-resolution spectra were obtained.  Solid
red lines represent data obtained with the Keck I telescope and HIRES.  
Long dashed blue lines represent data obtained with the VLT and UVES.  
Pink dashed lines represent data obtained with the HET and the HRS.  Green 
dash-dot lines represent data obtained at the McDonald Observatory 2.1-m 
telescope and the Sandiford Cassegrain Echelle spectrograph.  Orange 
dash-dot-dot-dot lines represent data obtained with the Magellan Clay 
telescope and the MIKE spectrometer. 
\label{Fig. 2}}
\end{figure}

\clearpage
\begin{figure}
\centering
\includegraphics{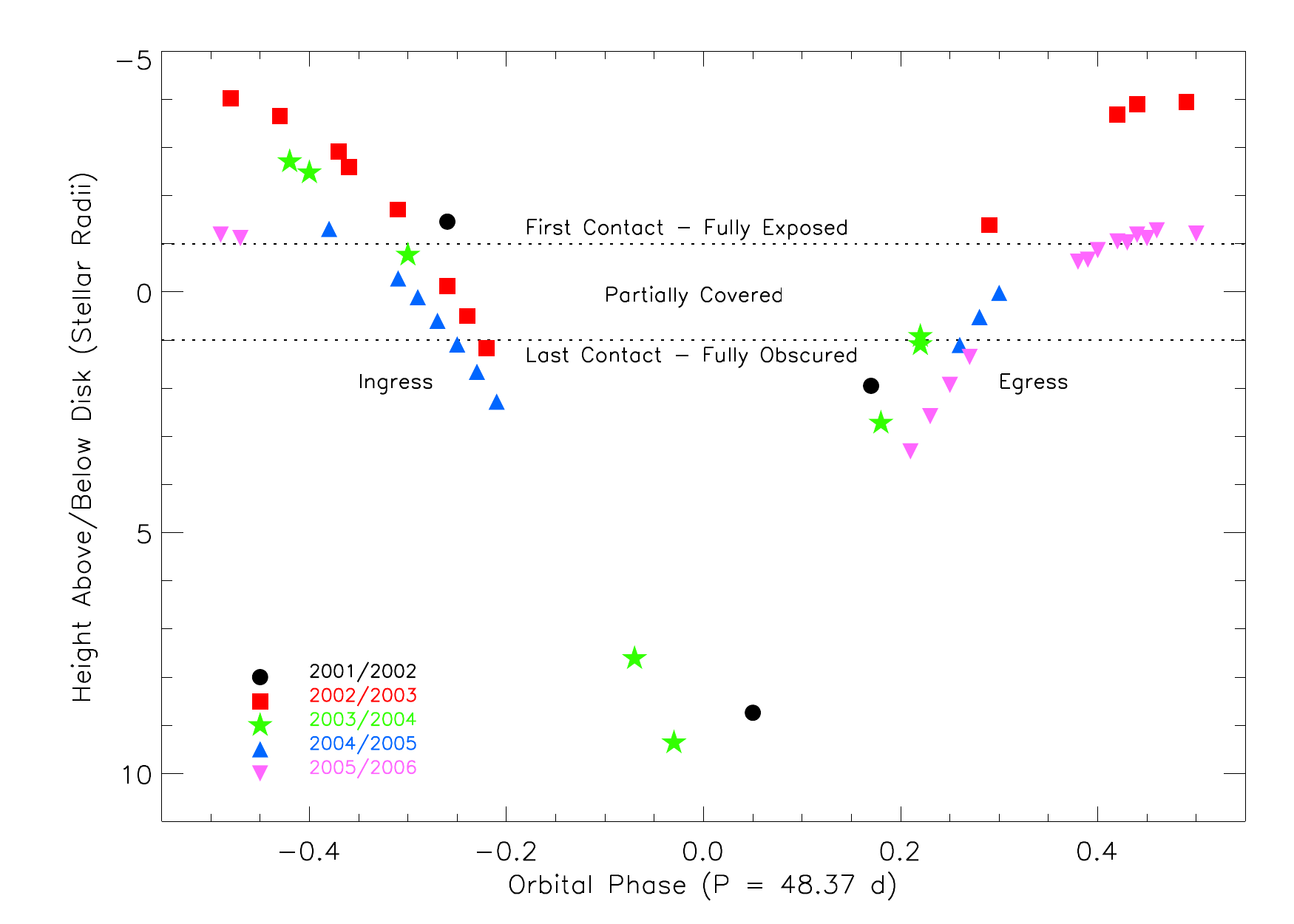}
\figcaption{Predicted height from W06 of star A above the occulting screen in stellar radii 
versus phase over the course of five different observing seasons.  Note that the same phase 
does not correspond to the same height above or below the disk due to the precession of 
the circumbinary ring.  The dotted lines in the figure represent the points at which the star 
makes first and last contact with the edge of the disk from our point of view.
\label{Fig. 3}}
\end{figure}

\clearpage
\begin{figure}
\centering
\includegraphics[width=6.5in,height=4in]{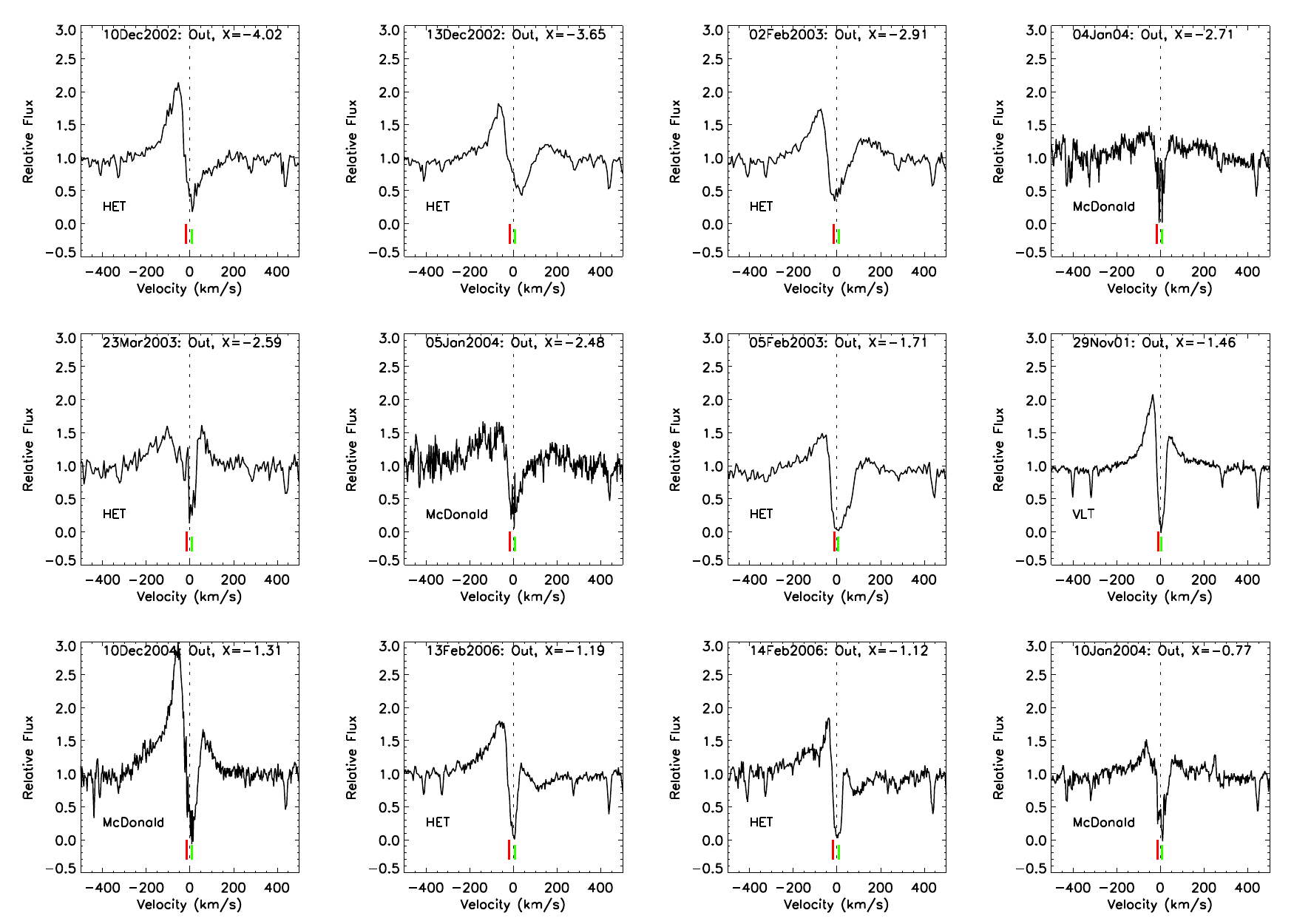}
\includegraphics[width=6.5in,height=4in]{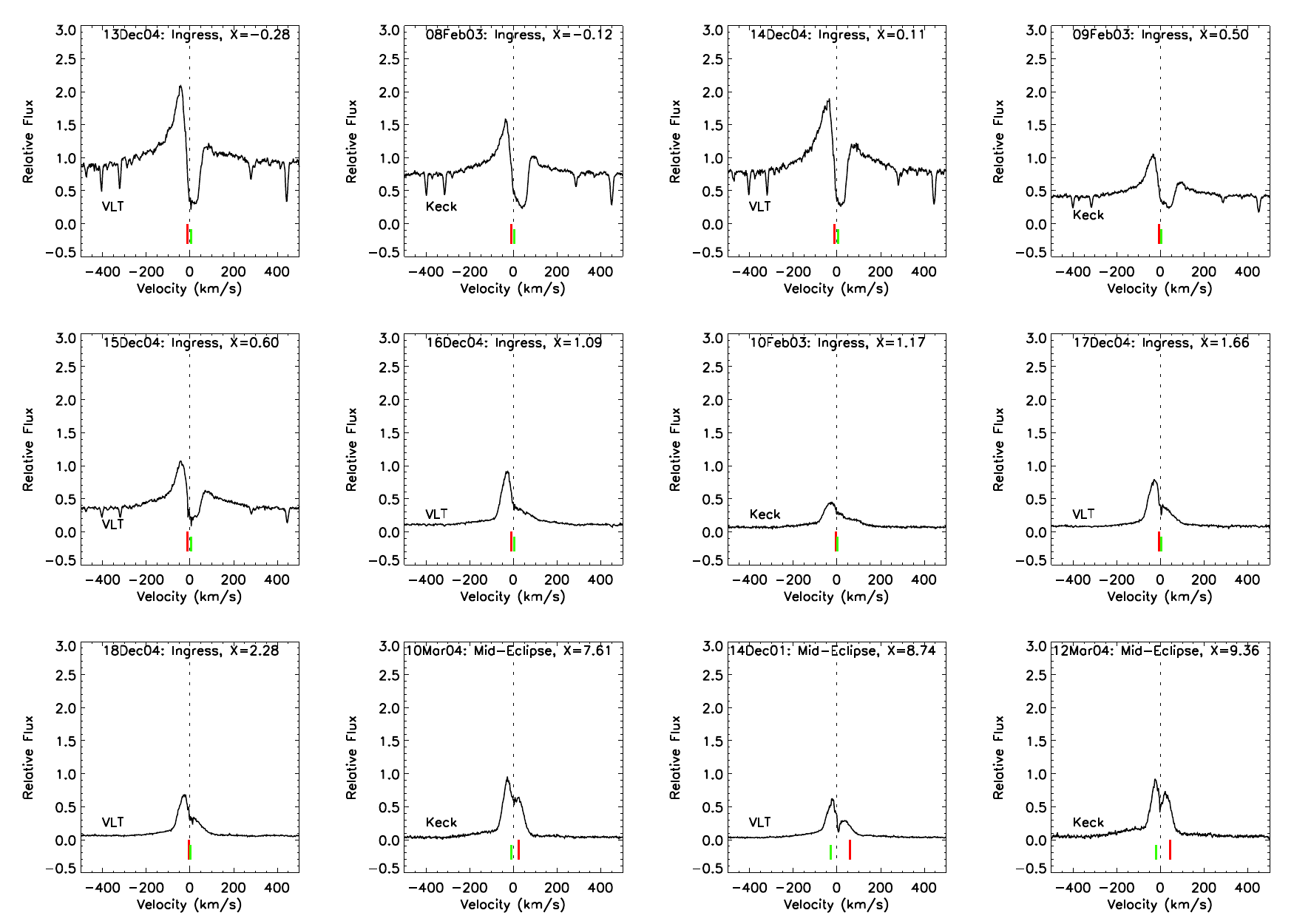}
\figcaption{From left to right, H$\alpha$ emission line profiles out-of-eclipse through ingress 
to mid-eclipse.  The red (long) and green (short) tickmarks represent the predicted velocities of 
stars A and B, respectively.  All profiles are shown in the reference frame of the system.
 \label{Fig. 4}}
\end{figure}

\clearpage
\begin{figure}
\centering
\includegraphics[width=6.5in,height=4in]{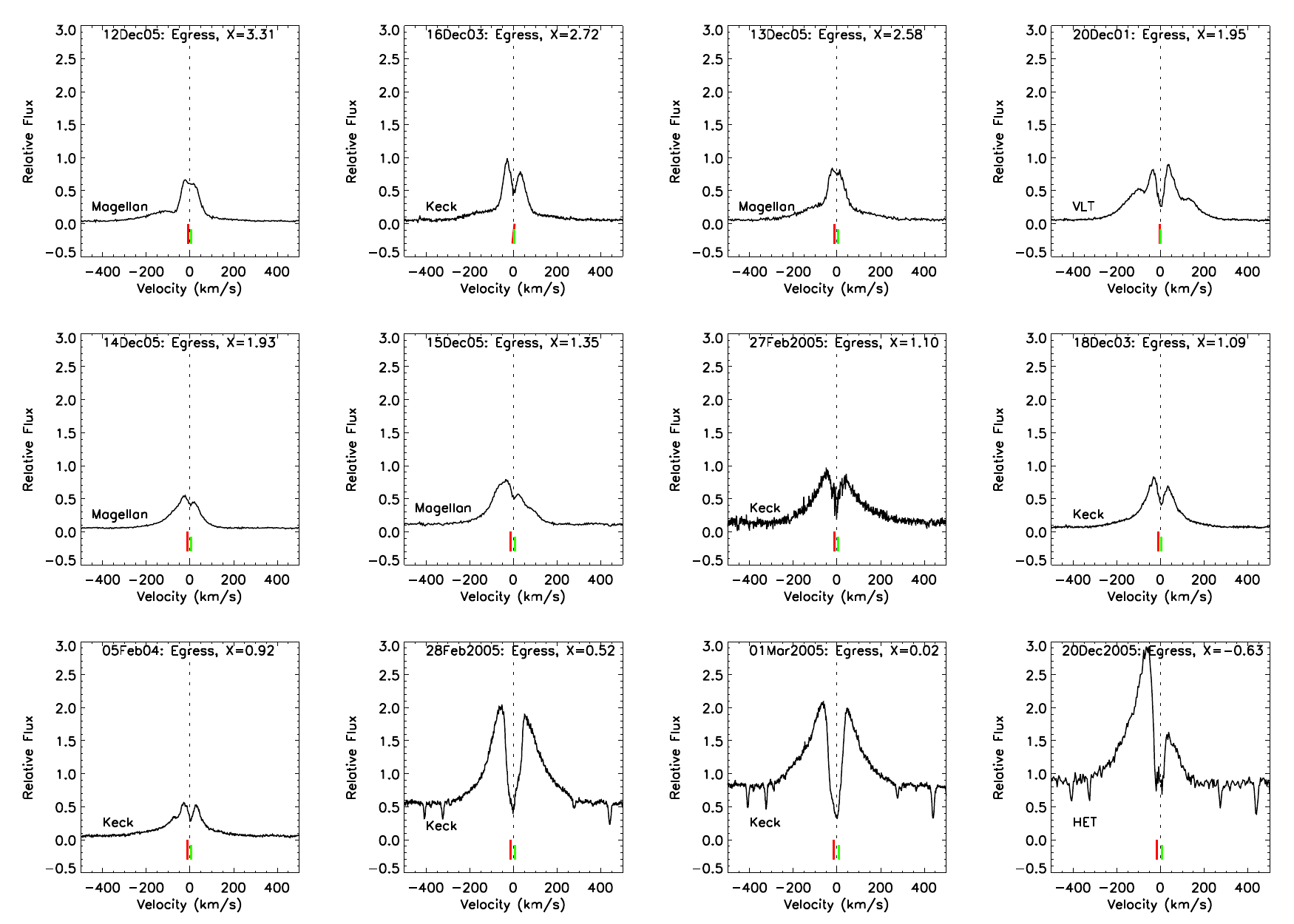}
\includegraphics[width=6.5in,height=4in]{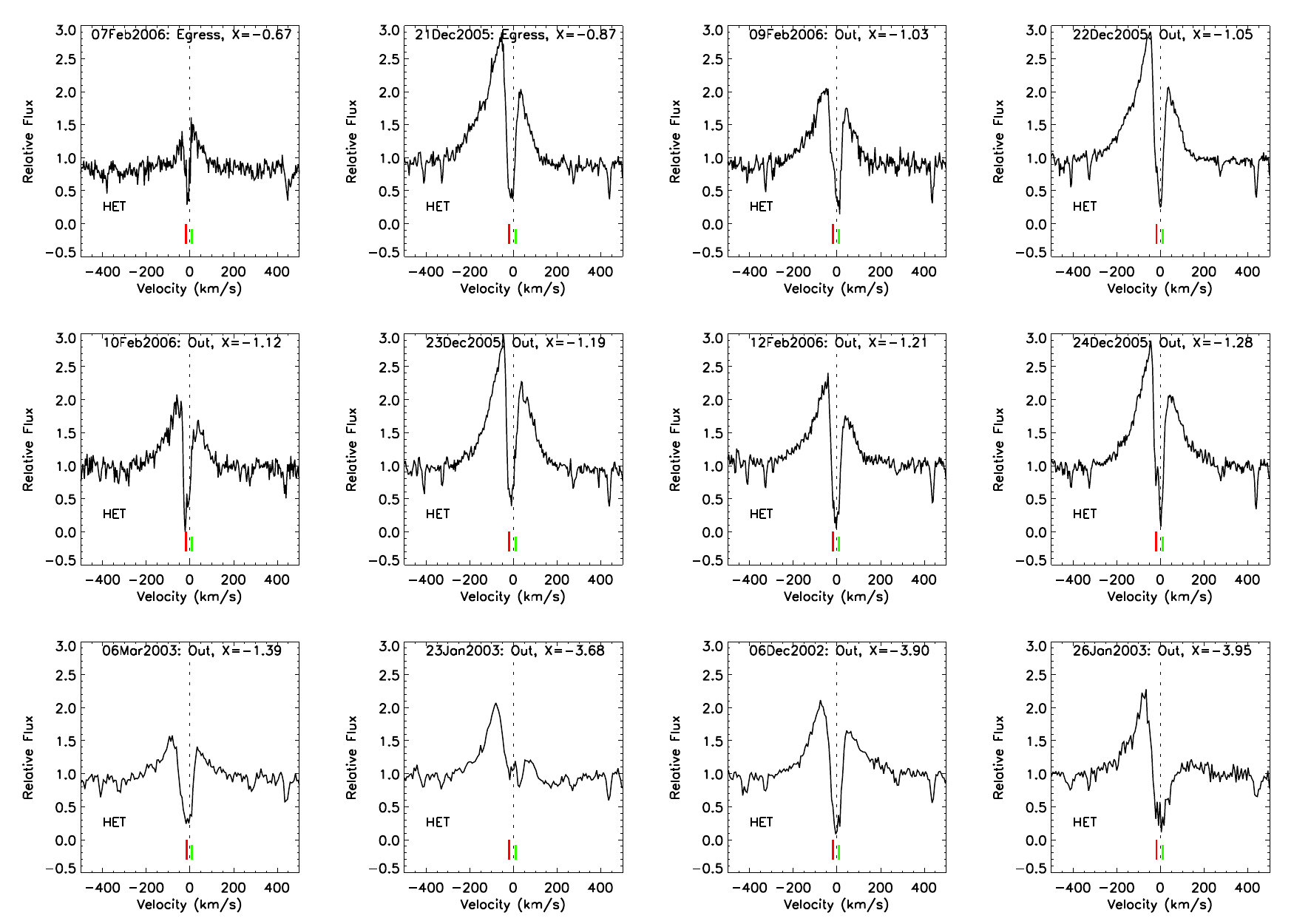}
\figcaption{Left to right, top to bottom, H$\alpha$ emission line profiles are shown during egress
and ending out-of-eclipse. The red (long) and green (short) tickmarks indicate the radial 
velocities of star A and B, respectively.  All profiles are shown in the reference frame of the system. 
\label{Fig. 5}}
\end{figure}

\clearpage
\begin{figure}[hbtp]
\centering
\includegraphics[width=6.5in,height=6.5in]{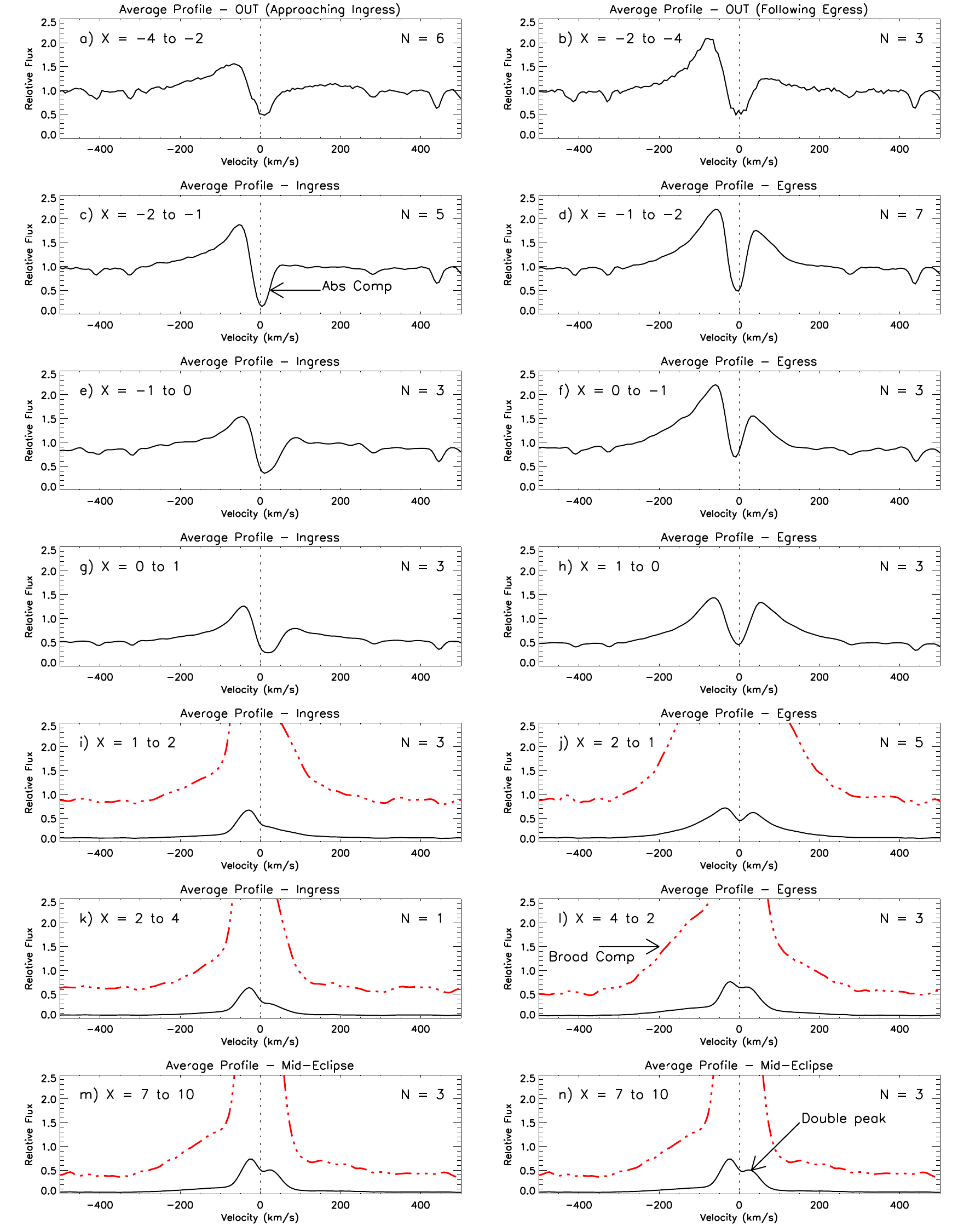}
\figcaption{The profiles from Figures 4 and 5 averaged according 
their distances above or below the occulting disk.  The letter N in 
the upper right hand corner designates how many spectra went into 
creating the average.  The bottom two panels are identical and represent
mid-eclipse.  Hence, the panels on the left side of the figure represent ingress, 
while the panels on the right side of the figure represent egress.  The red dash-dot-dot-dot
profile shown in the bottom six panels is the average profile for that distance bin
multiplied by 10 to show more clearly the blue wing present after the star is fully eclipsed.
\label{Fig. 6}}
\end{figure}

\clearpage
\begin{figure}[hbtp]
\centering
\includegraphics[totalheight=0.8\textheight]{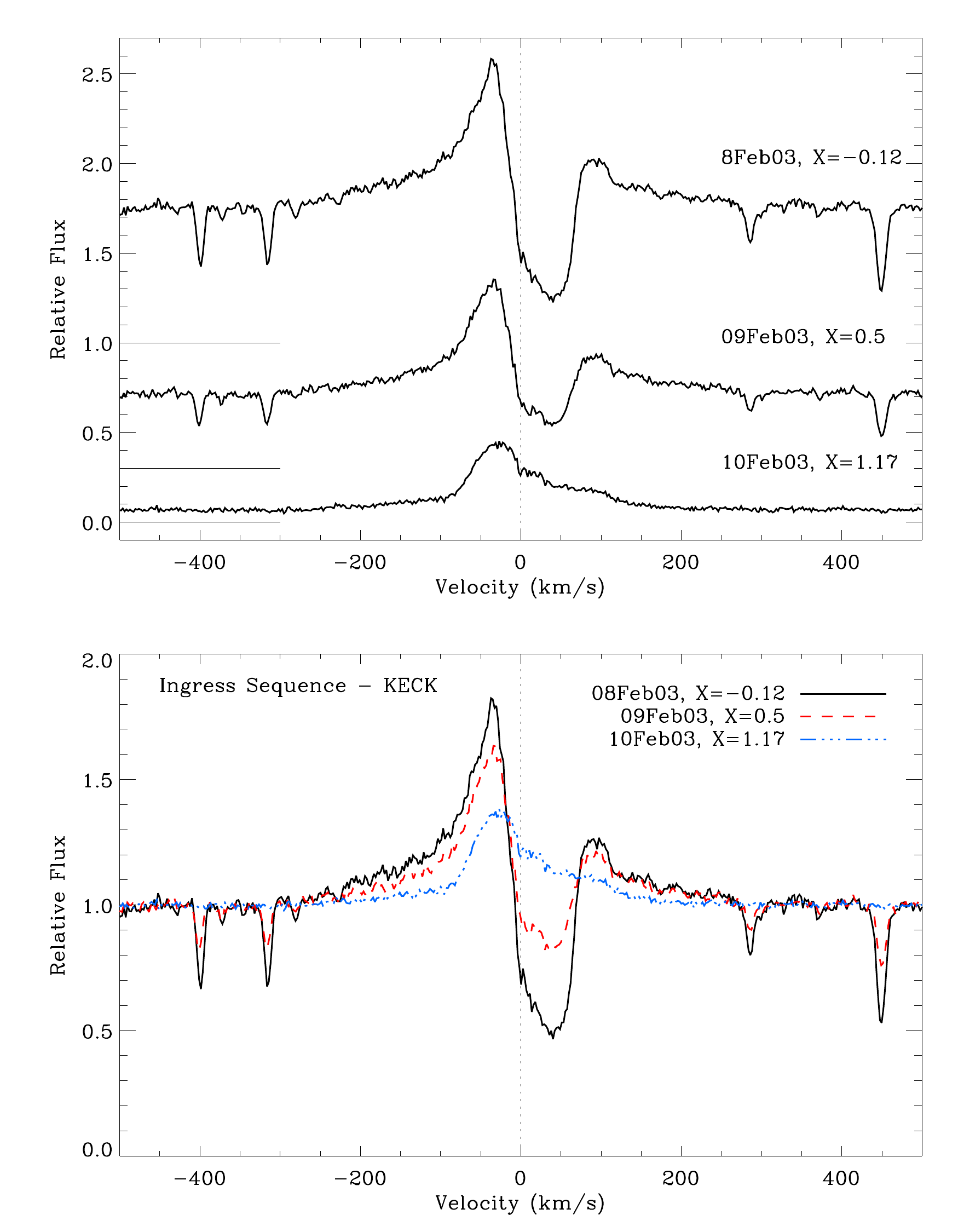}
\figcaption{The top panel shows a 3 day sequence of spectra obtained at Keck during 
an ingress in 2003 February.  Each spectrum has an offset applied, with the zero
point indicated by a small horizontal line along the y-axis.  The bottom panel shows
the spectra appropriately fluxed and normalized to 1.0 so that each spectrum can 
be plotted on top of one another.
\label{Fig. 7}}
\end{figure}

\clearpage
\begin{figure}[hbtp]
\centering
\includegraphics[totalheight=0.8\textheight]{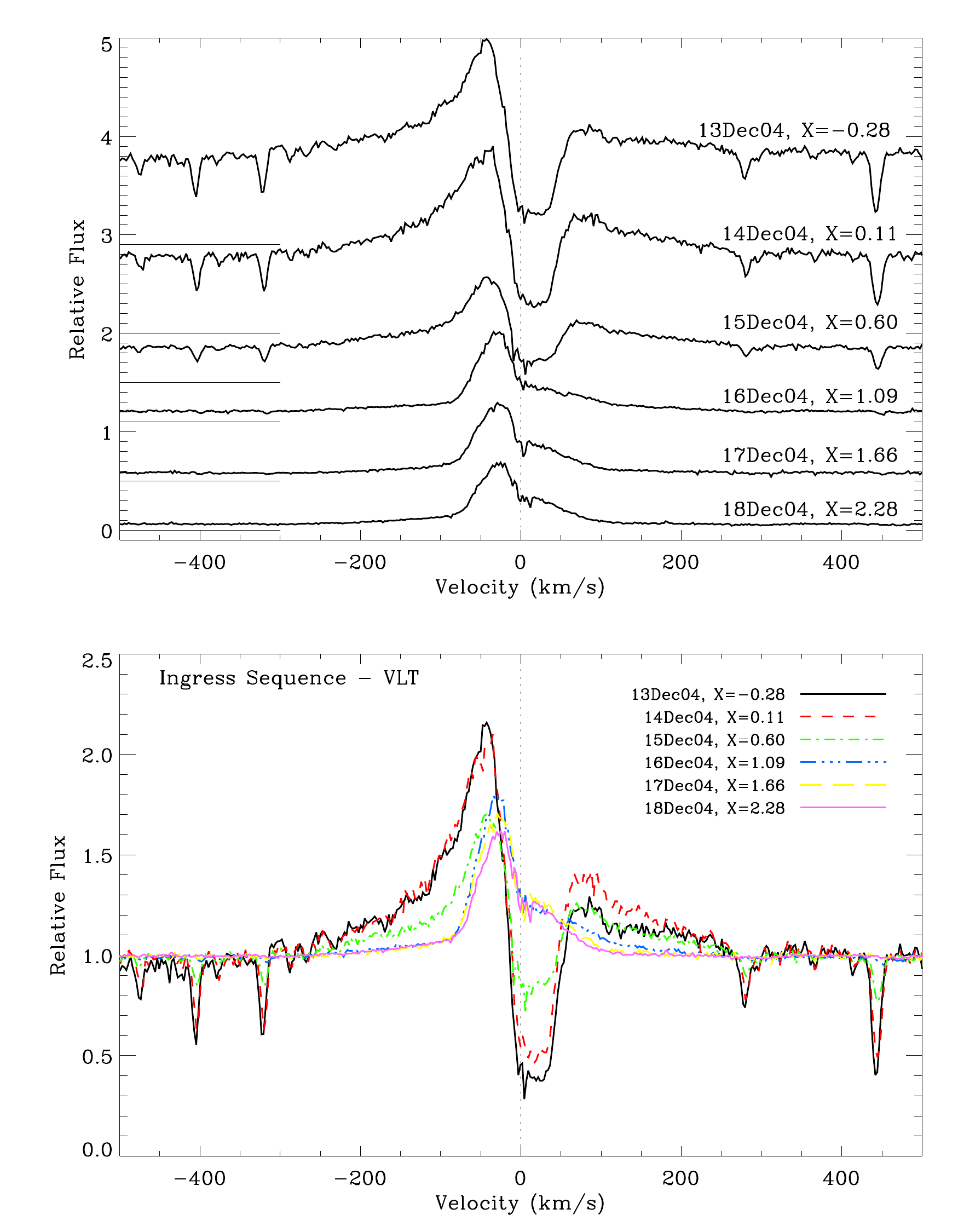}
\figcaption{The top panel shows a 6 day sequence of spectra obtained at the VLT during 
an ingress in 2004 December.  Each spectrum has an offset applied, with the zero
point indicated by a small horizontal line along the y-axis.  The bottom panel shows
the spectra appropriately fluxed and normalized to 1.0 so that each spectrum can 
be plotted on top of one another.
\label{Fig. 8}}
\end{figure}

\clearpage
\begin{figure}[hbtp]
\centering
\includegraphics[totalheight=0.8\textheight]{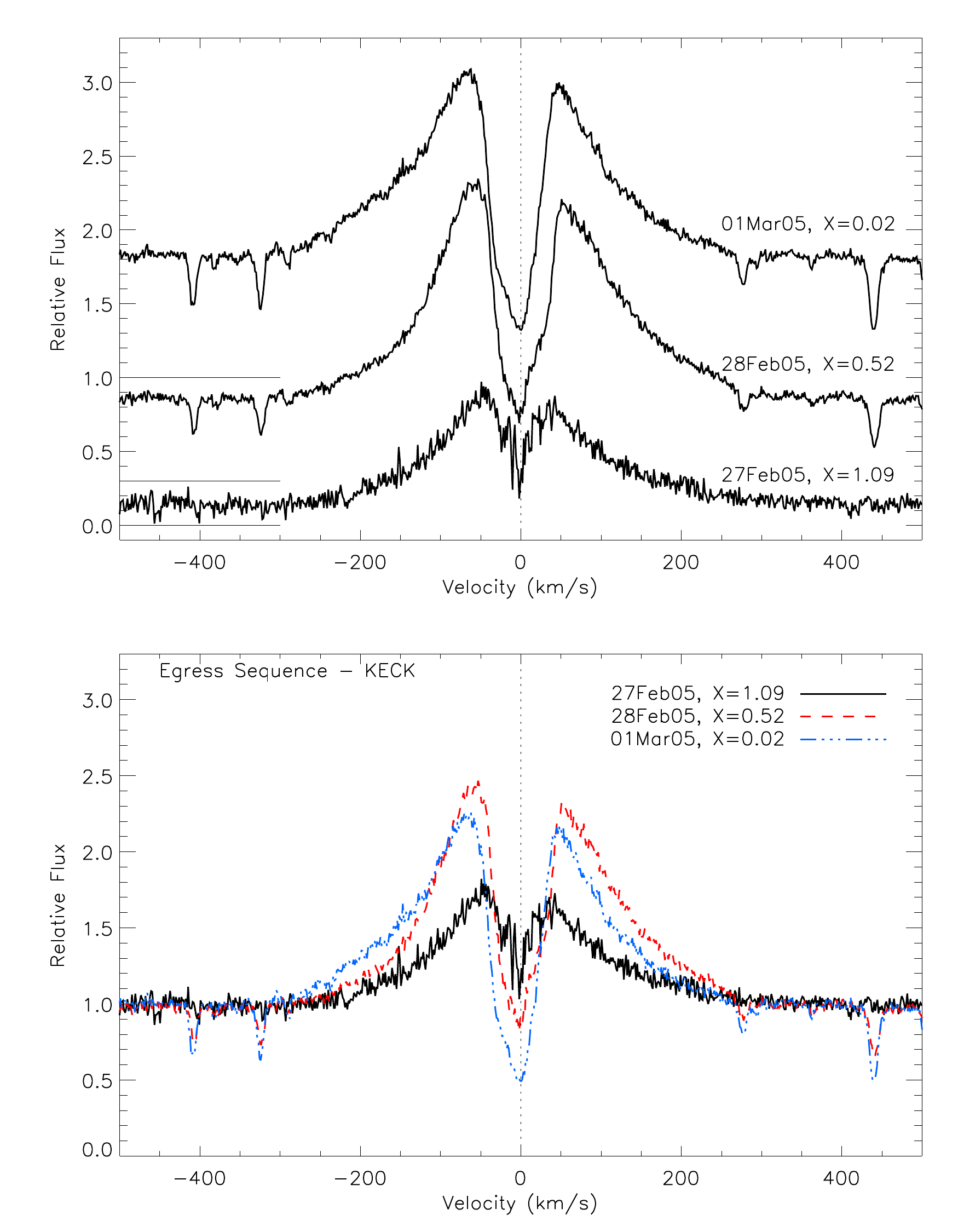}
\figcaption{The top panel shows three spectra obtained during egress.  Each
spectrum has an offset applied, with the zero point indicated by a small horizontal line
along y-axis.   The bottom panel shows the spectra appropriately fluxed and normalized 
to 1.0 so that each spectrum can be plotted on top of one another.
\label{Fig. 9}}
\end{figure}

\clearpage
\begin{figure}[hbtp]
\centering
\includegraphics[totalheight=0.8\textheight]{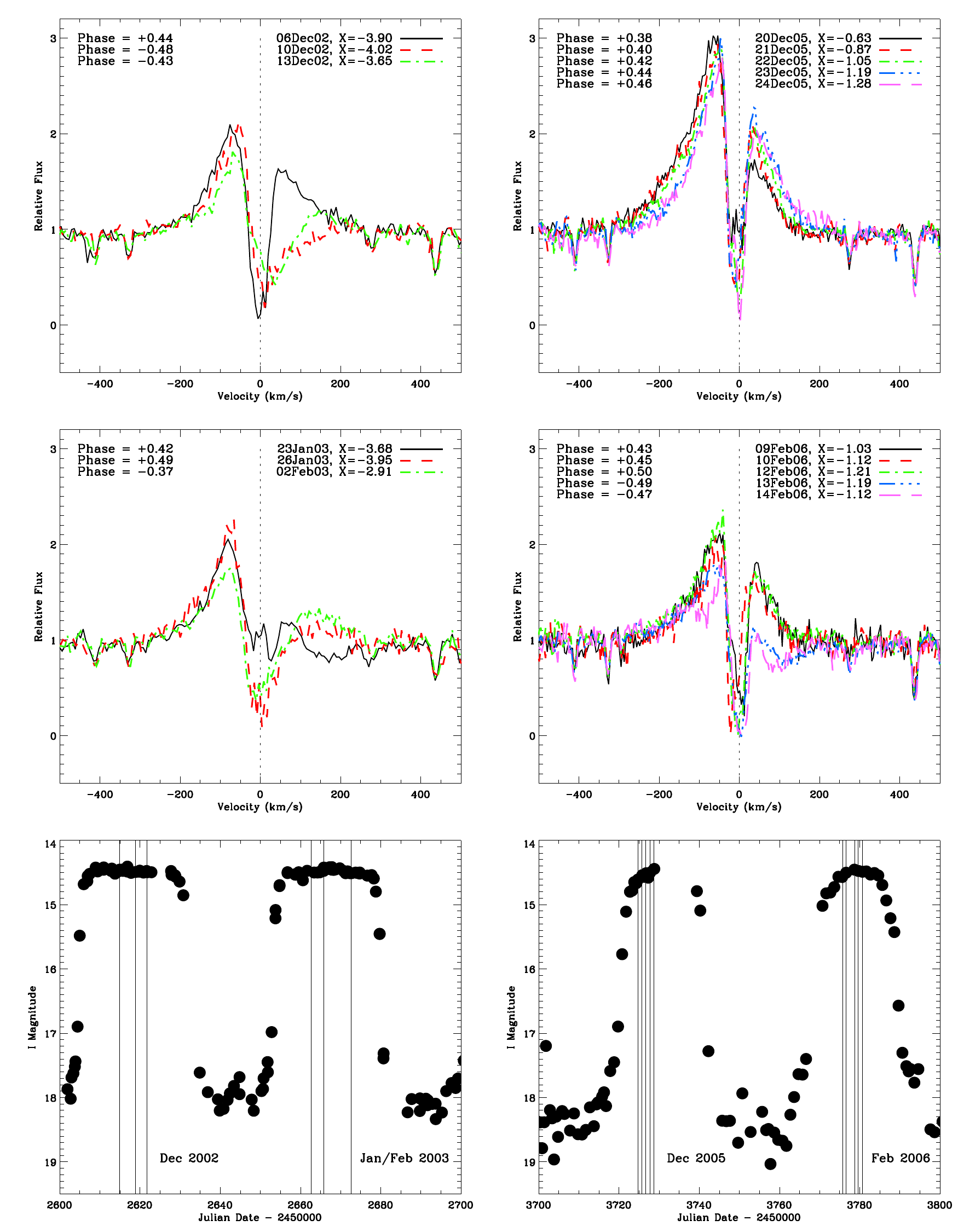}
\figcaption{H$\alpha$ profiles obtained while star A is the farthest out of eclipse.
The three left panels represent data taken during the 2002/2003 observing season.
The three panels on the right show data obtained during the 2005/2006 observing 
season.  
\label{Fig. 10}}
\end{figure}

\clearpage
\begin{figure}[hbtp]
\centering
\includegraphics{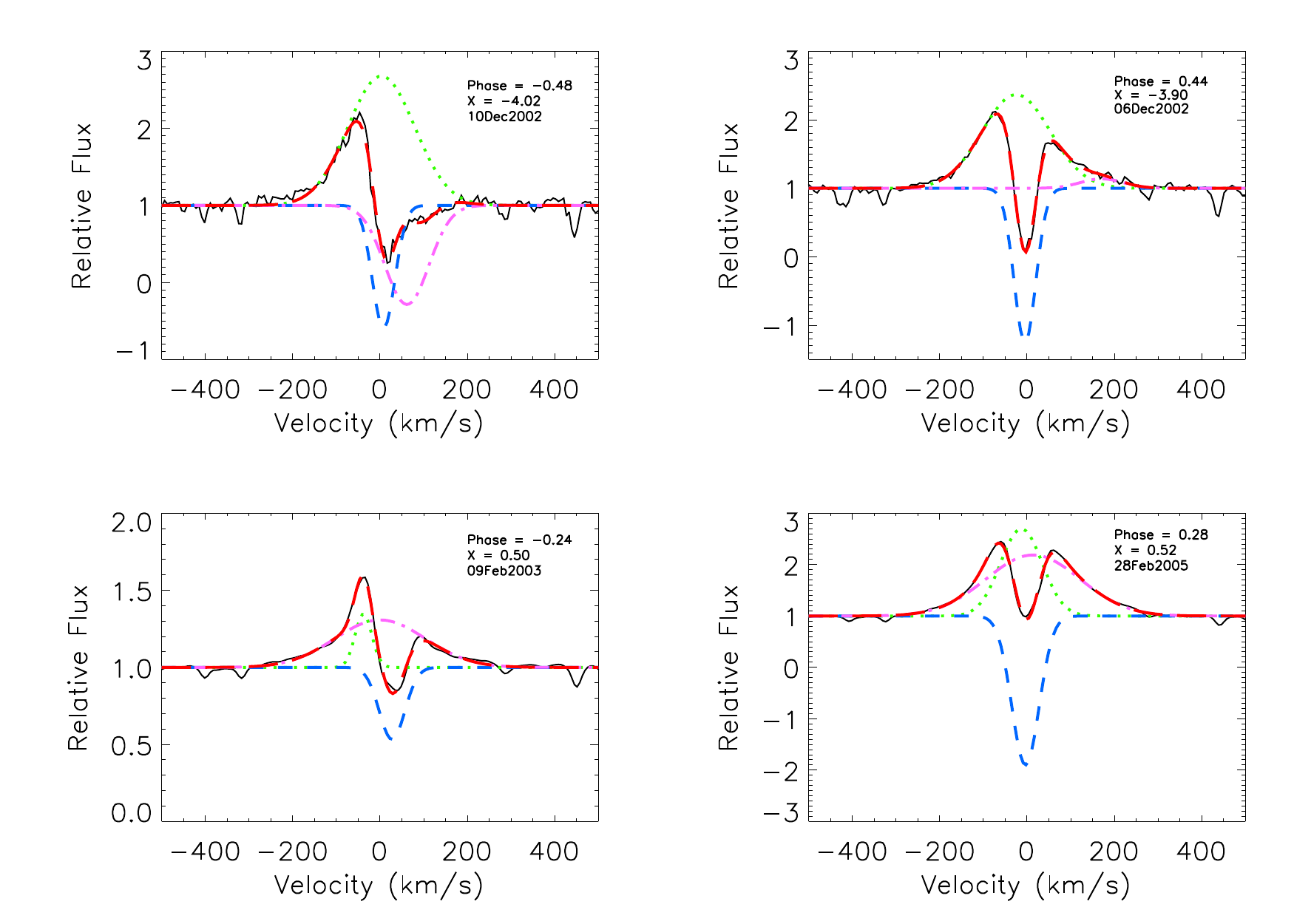}
\figcaption{Sample of profiles and three gaussian fits.  The combined fit is shown
in red.  
\label{Fig. 11}}
\end{figure}

\clearpage
\begin{figure}[hbtp]
\centering
\includegraphics[totalheight=0.8\textheight]{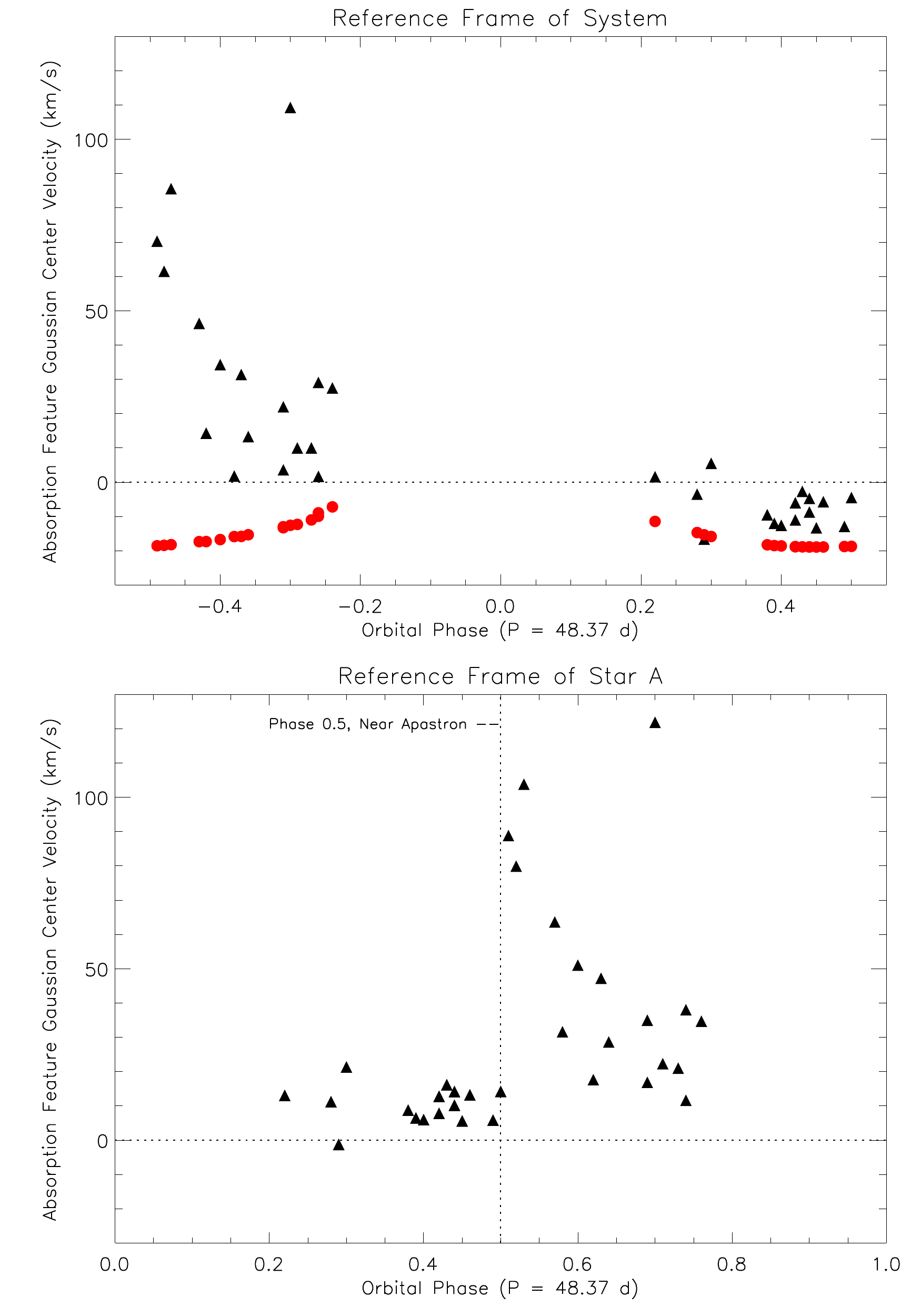}
\figcaption{The black triangles represent the velocity center of the gaussian fit to the
absorption component of the observed profiles.  Plotted in red are the predicted velocities
of Star A for each observation.  The horizontal dotted line represents the rest frame of the
system.  The bottom panel shows the velocity of the absorption 
component versus phase in the reference frame of star A as viewed from the Earth, and
the horizontal dotted line represents the rest frame of the star.  
\label{Fig. 12}}
\end{figure}

\clearpage
\begin{figure}[hbtp]
\centering
\includegraphics[totalheight=0.8\textheight]{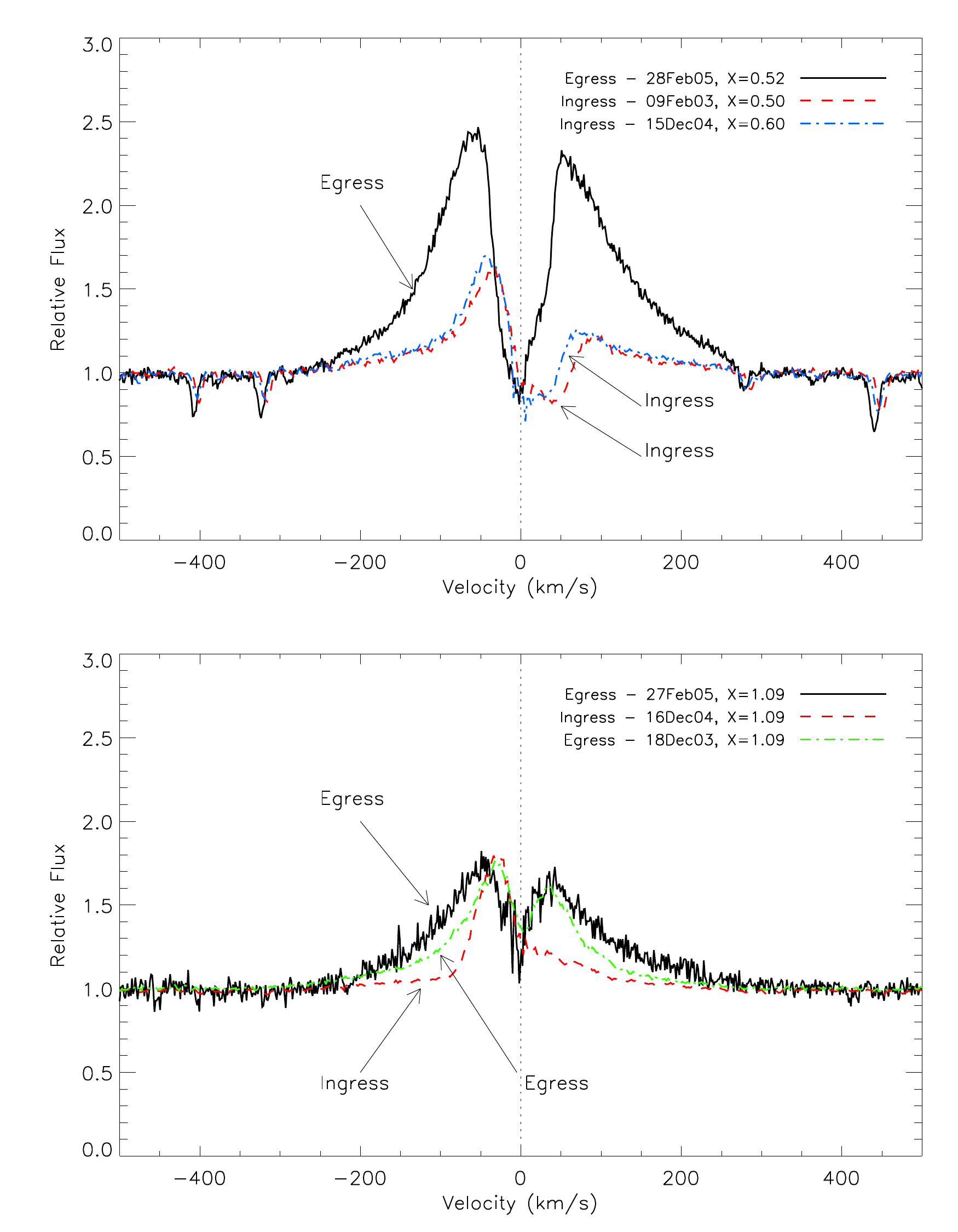}
\figcaption{The top panel shows a comparison of ingress and egress spectra when the
star was at a distance below the CBD edge of $\Delta$X $\sim$ 0.5.
The bottom panel shows the same comparison, but for spectra taken when the star
was at the edge of the CBD but still occulted. 
\label{Fig. 13}}
\end{figure}

\clearpage
\begin{figure}[hbtp]
\centering
\includegraphics[totalheight=0.8\textheight]{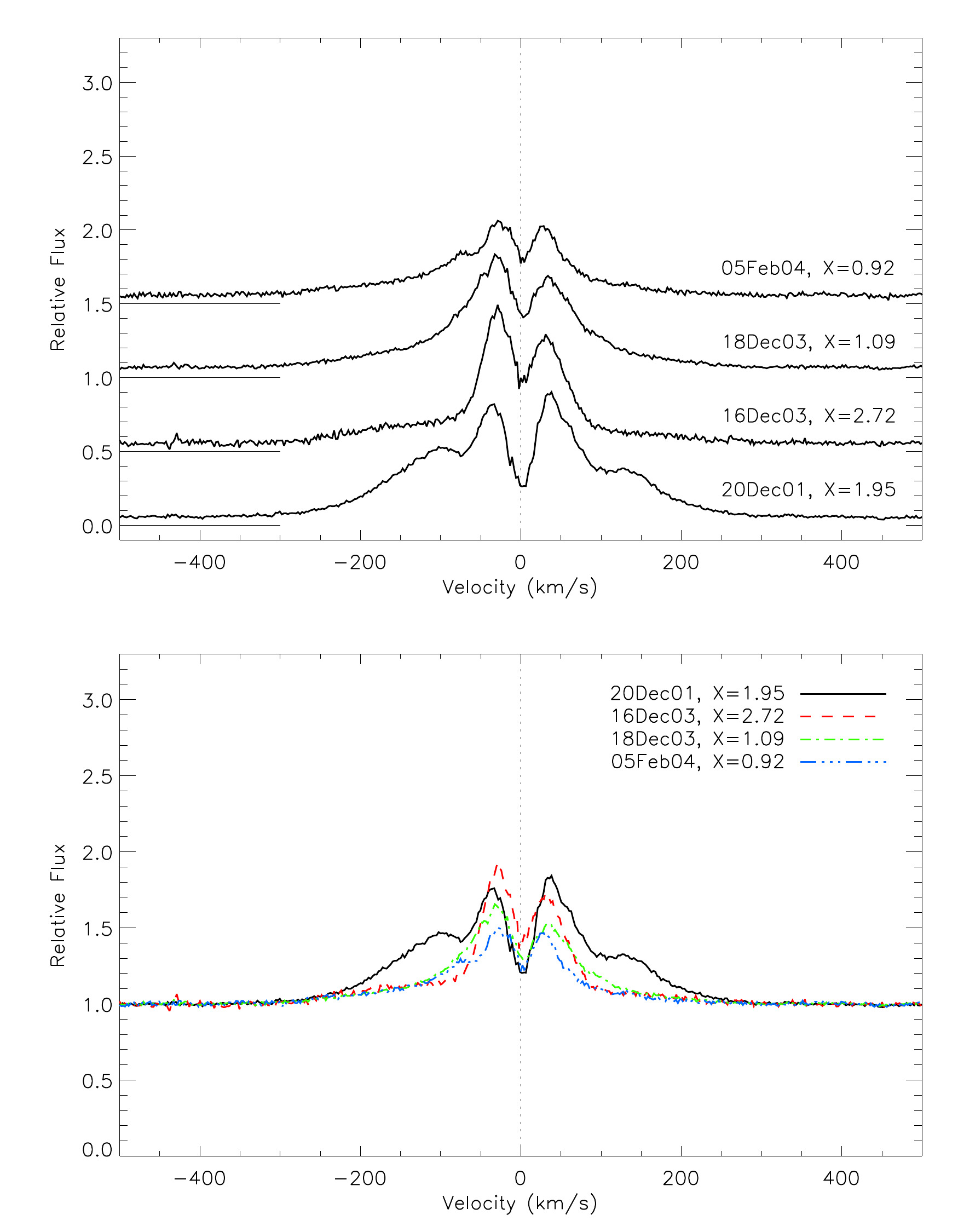}
\figcaption{The top panel shows 4 spectra obtained during various egresses over the
course of 2 observing seasons (2001/2001, 2002/2003).  Each spectrum has an offset 
applied, with the zero point indicated by a small horizontal line along the y-axis.  
The bottom panel shows the spectra appropriately fluxed and normalized to 1.0 so that 
each spectrum can be plotted on top of one another. 
\label{Fig. 14}}
\end{figure}

\clearpage
\begin{figure}[hbtp]
\centering
\includegraphics[totalheight=0.8\textheight]{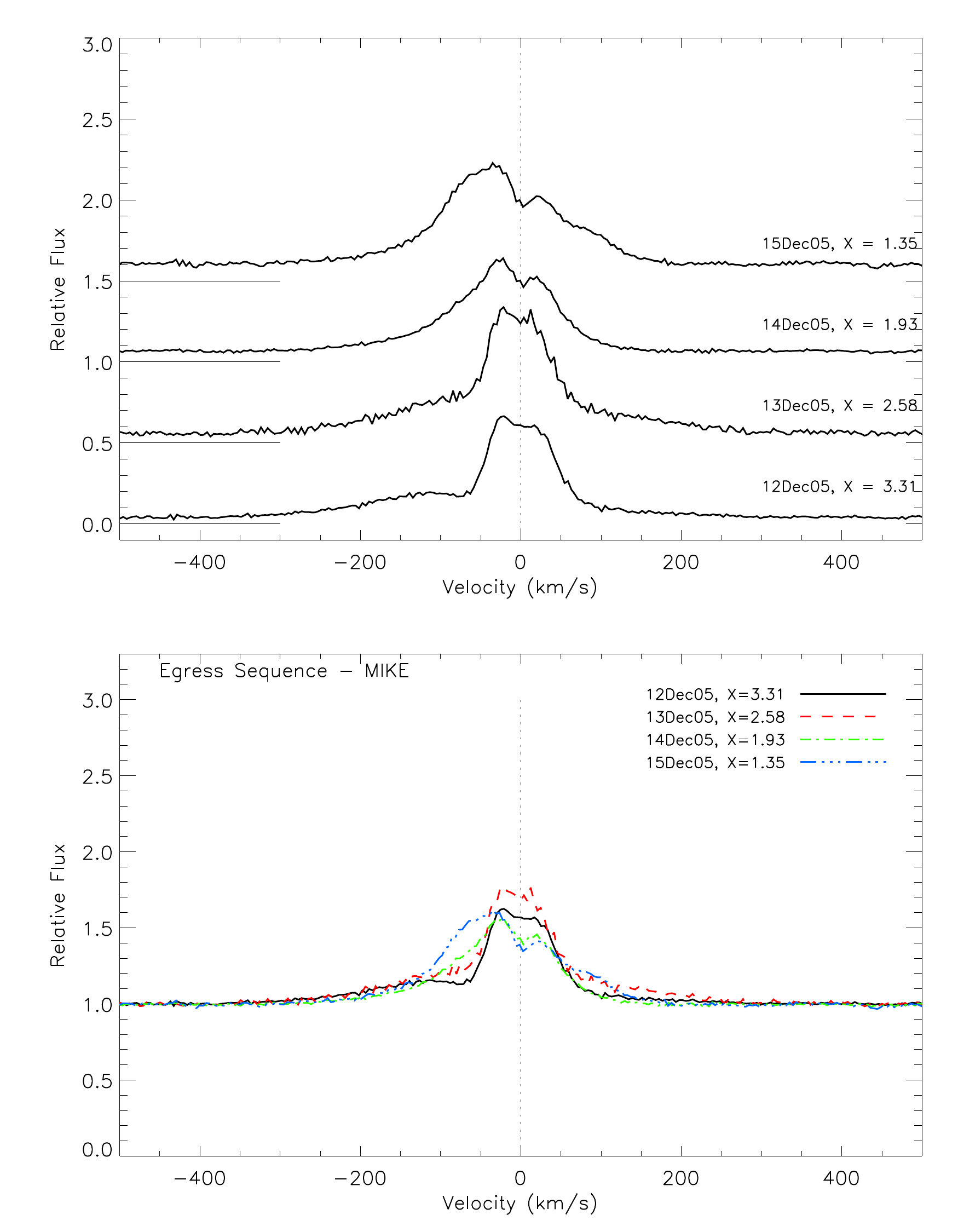}
\figcaption{The top panel shows 4 spectra obtained on consecutive nights during an egress
that took place in 2005 December.  Each spectrum has an offset applied, with the zero point 
indicated by a small horizontal line along the y-axis.  The bottom panel shows the spectra 
appropriately fluxed and normalized to 1.0 so that each spectrum can be plotted on top of one 
another. 
\label{Fig. 15}}
\end{figure}

\clearpage
\begin{figure}[hbtp]
\centering
\includegraphics[totalheight=0.8\textheight]{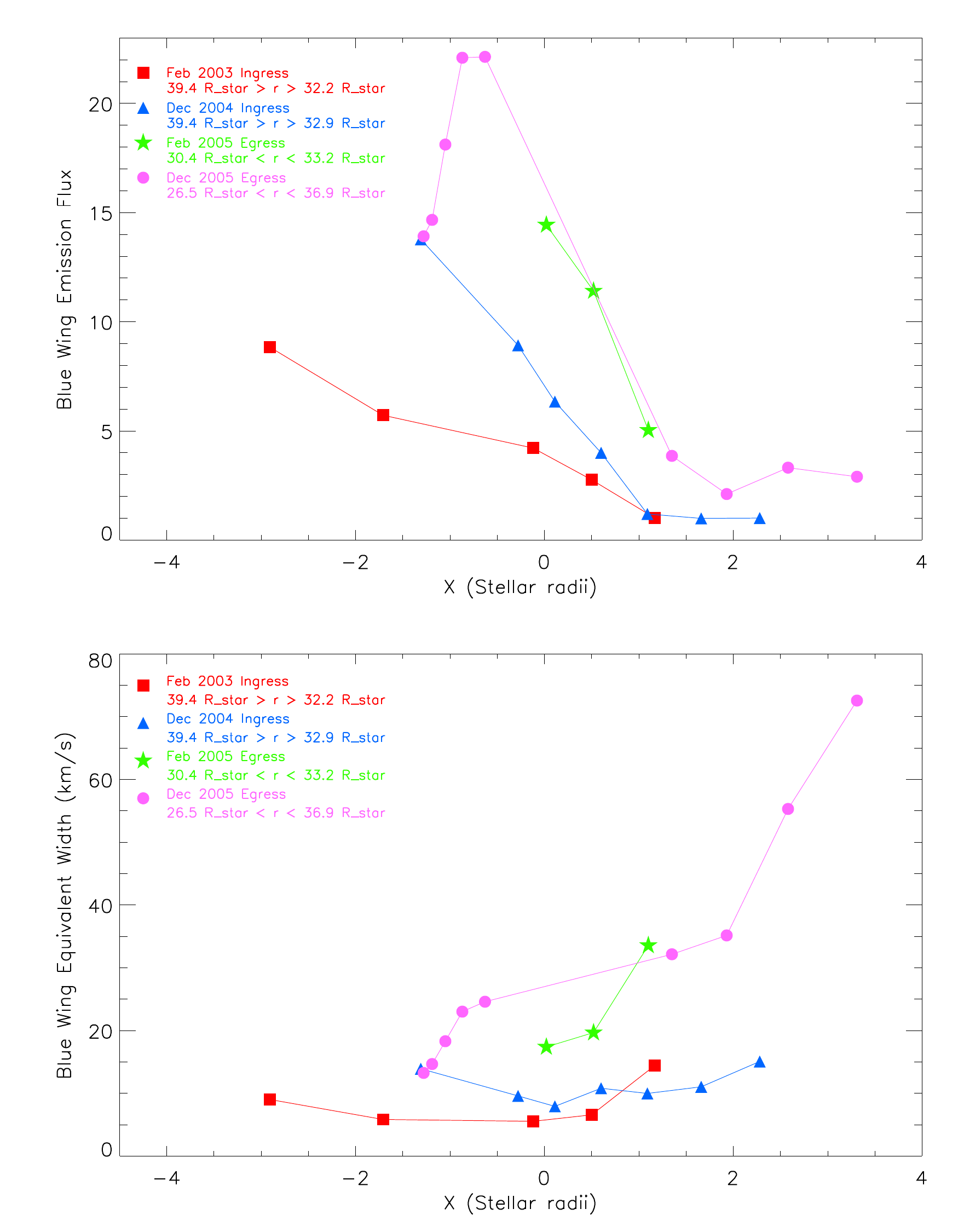}
\figcaption{The top panel shows the blue wing emission flux between --85 and --280 
km s$^{-1}$ measured for star A as it undergoes various ingress/egress events as 
noted in the figure.  The bottom panel shows the equivalent width of the blue wing
between --85 and --280 km s$^{-1}$ for the same set of ingress/egress events.
\label{Fig. 16}}
\end{figure}

\clearpage
\begin{figure}[hbtp]
\centering
\includegraphics[totalheight=0.8\textheight]{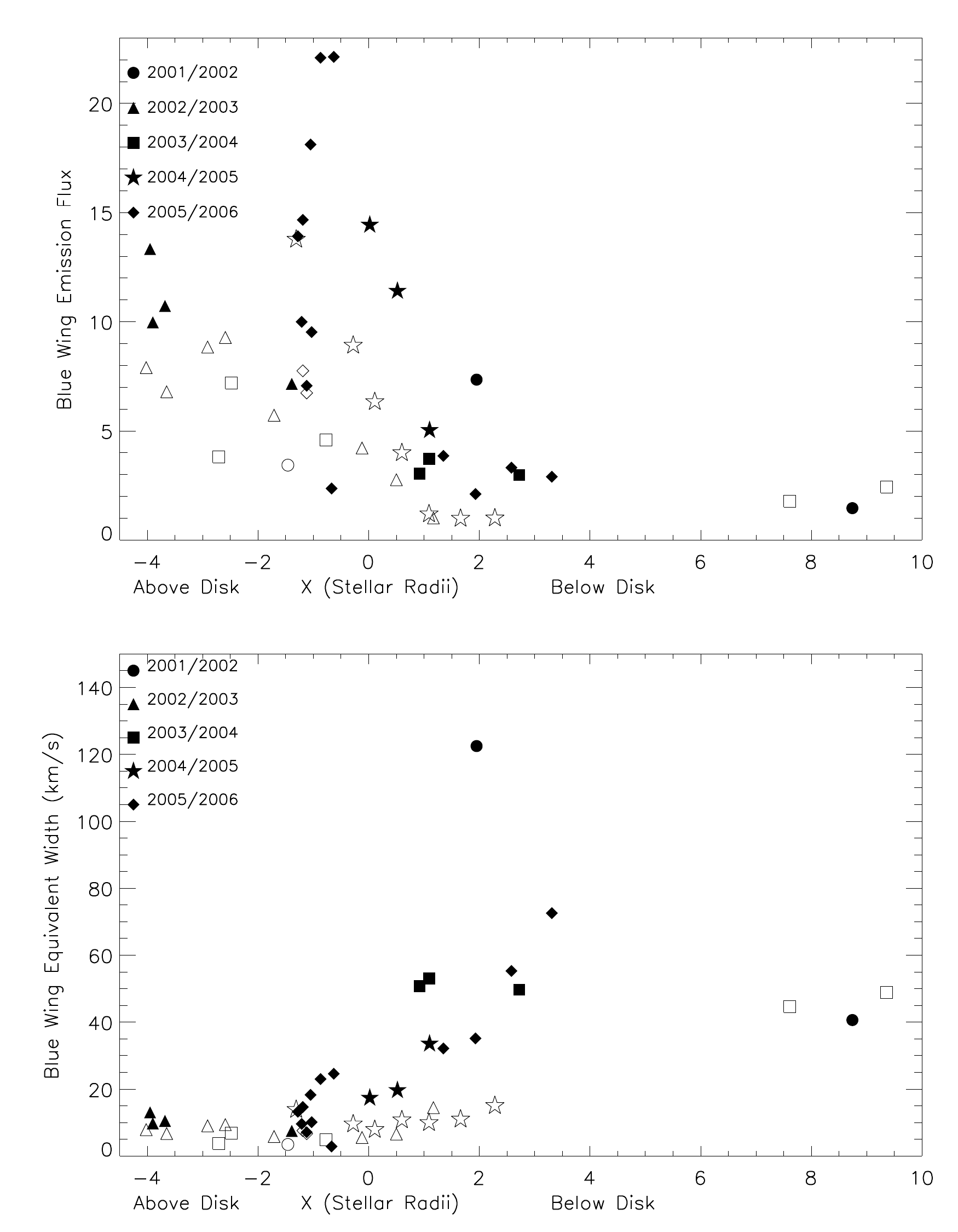}
\figcaption{The top panel shows the blue wing emission flux between --85 and --280 
km s$^{-1}$ measured for star A for all 48 observations.  The bottom panel shows the 
equivalent width of the blue wing between --85 and --280 km s$^{-1}$ for the same set of 
observations.
\label{Fig. 17}}
\end{figure}

\clearpage
\begin{figure}[hbtp]
\centering
\includegraphics{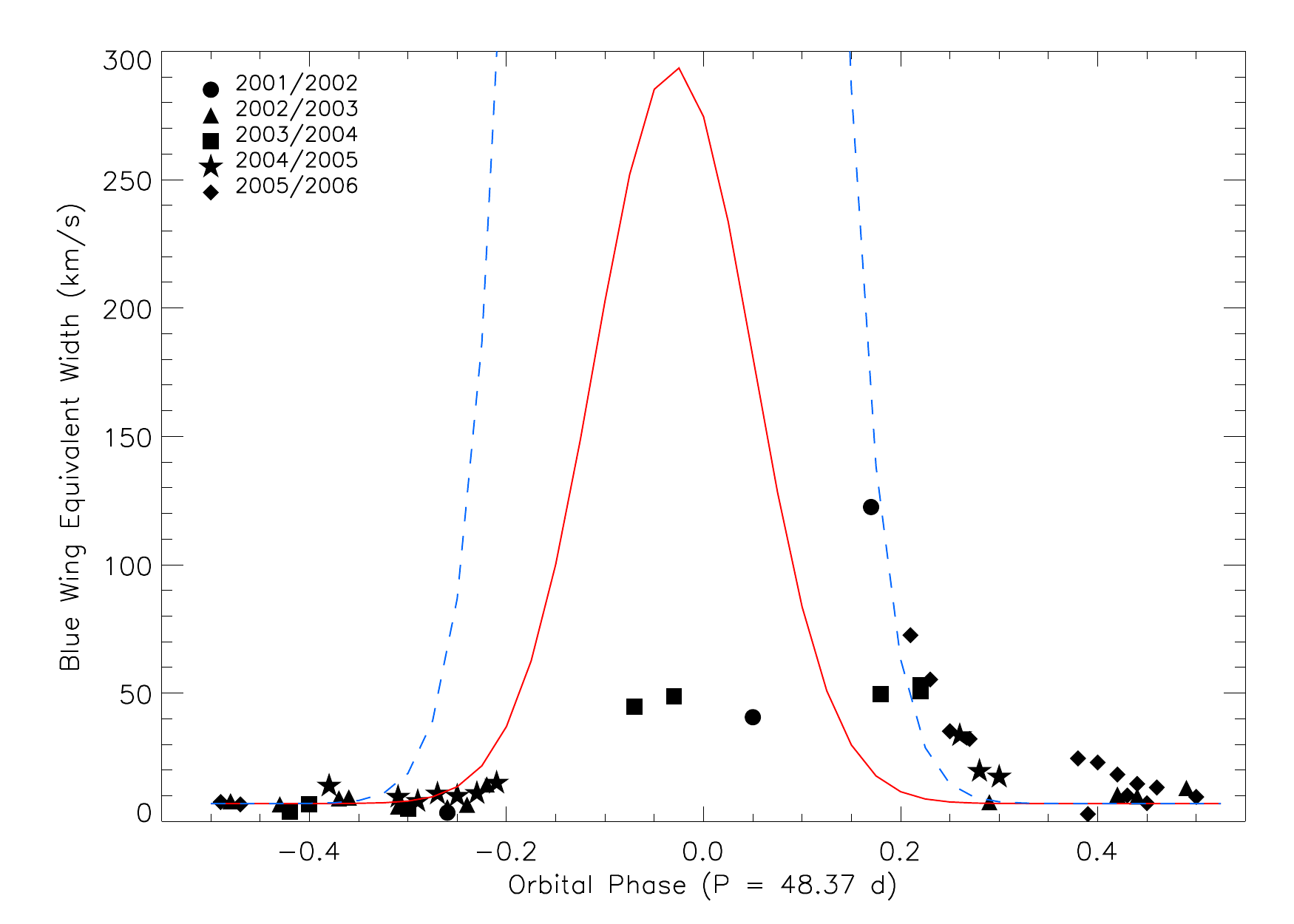}
\figcaption{The EW of the H$\alpha$ emission between  --85 and --280 
km s$^{-1}$.  The solid red line represents a gaussian fit to match the gas accretion
rate from Figure 2 of Artymowicz \& Lubow (1996).  The dashed blue line is the same
gaussian scaled to match the egress EW.
\label{Fig. 18}}
\end{figure}

\end{document}